\documentclass[11pt]{article}
\usepackage{amsmath}
\usepackage{latexsym}
\usepackage{epsf}
\def\IR{\relax{\rm I\kern-.18em R}}
\def\I1{\relax{\rm 1\kern-.40em 1}}
\def\IZ{\relax{\rm Z\kern-.40em Z}}
\def\be{\begin{equation}}
\def\ee{\end{equation}} 
\def\Dsl{\relax{\rm D \kern-.50em /}}
\begin{document}
\thispagestyle{empty}
\begin{flushright} LPTENS-13/14 \end{flushright}
\vskip 1cm
\begin{center}
{\bf \Large Introduction to the STANDARD MODEL\\
\vskip 0.3cm
of the Electro-Weak Interactions }\\
\vskip 1cm

{\Large JOHN ILIOPOULOS}
\vskip 1cm

Laboratoire de Physique Th\'eorique \\ de L'Ecole Normale Sup\'erieure \\
75231 Paris Cedex 05, France

\vskip 5cm

Lectures given at the 2012 CERN Summer School
\vskip 1cm
June 2012, Angers, France

\end{center}

\newpage

\section{Introduction}

These are the notes of a set of four lectures which I gave at the 2012 CERN Summer
School of Particle Physics. They were supposed to serve as introductory
material to more specialised lectures. The students were mainly young graduate
students in Experimental High Energy Physics. They were supposed to be
familiar with the phenomenology of Particle Physics and to have a working
knowledge of Quantum Field Theory and the techniques of Feynman
diagrams. The lectures were concentrated on the physical ideas underlying the
concept of gauge invariance, the mechanism of spontaneous symmetry breaking
and the construction of the Standard Model. Although the methods of computing
higher order corrections and the theory of
renormalisation were not discussed at all in the lectures, the general concept
of renormalisable versus non-renormalisable theories was supposed to be
known. 

The plan of the notes follows that of the lectures with five sections: 

$\bullet$ A brief summary of the phenomenology of the electromagnetic and the
weak interactions.

$\bullet$ Gauge theories, Abelian and non-Abelian.

$\bullet$ Spontaneous symmetry breaking.

$\bullet$ The step-by-step construction of the Standard Model.

$\bullet$ The Standard Model and experiment.

It is only in the last part that the notes differ from the actual lectures,
because I took into account the recent evidence for a Higgs boson. 
\vskip 0.5cm

It is generally admitted that progress in physics occurs when an unexpected
experimental result contradicts the established theoretical beliefs. As
Feynman has put it ``progress in Physics is to prove yourself wrong as soon as
possible.'' This has been the rule in the past, but there are exceptions. The
construction of the Standard Model is one of them. In the late sixties weak
interactions were well described by the Fermi current x current theory and
there was no compelling experimental reason to want to change it. Its problems
were theoretical. It was only a phenomenological model which, in the technical
language, was non-renormalisable. In practice this means that any attempt to
compute higher order corrections in the standard perturbation theory would give meaningless divergent results. So, the motivation was aesthetic rather than experimental, it was the search of mathematical consistency and theoretical elegance. In fact, at the beginning, the data did not seem to support the theoretical speculations. Although the history of these ideas is a fascinating subject, I decided not to follow the historical evolution. The opposite would have taken more than four lectures to develop. I start instead from the experimental data known at present and show that they point unmistakably to what is known as the Standard Model. It is only at the last section where I recall its many experimental successes.

\section{Phenomenology of the Electro-Weak interactions: A reminder}
\subsection{The Elementary Particles}

The notion of an ``Elementary Particle'' is not well-defined in High Energy
Physics. It evolves with time following the progress in the experimental
techniques which, by constantly increasing the resolution power of our
observations, have shown that systems which were believed to be
``elementary'', are in fact composite out of smaller constituents. So, in the
last century we went through the chain:

molecules $\rightarrow $ atoms $\rightarrow $ electrons + nuclei $\rightarrow
$ electrons + protons + neutrons $\rightarrow $ electrons + quarks
$\rightarrow $ ???

There is no reason to believe that there is an end in this series and, even
less, that this end has already been reached. Table 1 summarises
our present knowledge.

\begin{center}
\begin{table}
\begin{tabular}{|c|c|c|}  \hline\hline
\multicolumn{3}{|c|}{\bf{TABLE OF ELEMENTARY PARTICLES}} \\ \hline\hline
\multicolumn{3}{|c|}{QUANTA OF RADIATION} \\ \hline
\multicolumn{2}{|c|}{Strong Interactions} & 
Eight gluons \\ \hline
\multicolumn{2}{|c|}{Electromagnetic Interactions} & 
Photon ($\gamma$) \\ \hline 
\multicolumn{2}{|c|}{Weak Interactions} & 
Bosons $W^+$ , $W^-$ , $Z^0$ \\ \hline
\multicolumn{2}{|c|}{Gravitational Interactions} & 
Graviton (?) \\ \hline
\multicolumn{3}{|c|}{MATTER PARTICLES} \\ \hline
  & 
Leptons  & Quarks \\ \hline
1st Family & 
${\nu}_e$ , $e^-$  & $u_a$ , $d_a$ , $a=1,2,3$ \\ \hline
2nd Family  & 
${\nu}_{\mu}$ , ${\mu}^-$ & $c_a$ , $s_a$ , $a=1,2,3$ \\ \hline
3rd Family & 
${\nu}_{\tau}$ , ${\tau}^-$ & $t_a$ , $b_a$ , $a=1,2,3$ \\ \hline
\multicolumn{3}{|c|}{HIGGS BOSON } \\ \hline\hline
\end{tabular}
\caption{This Table shows our present ideas on the 
structure of matter. Quarks and gluons do not exist as
free particles and the graviton has not yet been observed.}
\label{TabElPart}
\end{table}
\end{center}

Some remarks concerning this Table:

$\bullet$ All interactions are produced by the exchange of virtual quanta. For the
strong, electromagnetic and weak interactions they are vector (spin-one) fields, while
the graviton is assumed to be a tensor, spin-two field. We shall see in these
lectures that this property is well understood in the framework of gauge
theories.

$\bullet$ The constituents of matter appear to be all spin one-half particles. They are
divided into  quarks, which are hadrons, and ``leptons'' which have no strong
interactions. No deep explanation is known neither for their number, (why
three families?), nor for their properties, such as their quantum numbers. We
shall come back to this point when we discuss the gauge theory models. In the
framework of some theories going beyond the Standard Model, 
such as supersymmetric theories, among the matter constituents we can find
particles of zero spin. 

$\bullet$ Each quark species, called ``flavour'', appears under three forms, often called ``colours'' (no
relation with the ordinary sense of the words). 

$\bullet$ Quarks and gluons do not appear as free particles. They form a large
number of bound states, the hadrons. This property of ``confinement'' is one
of the deep unsolved problems in Particle Physics.

$\bullet$ Quarks and leptons seem to fall into three distinct groups, or
``families''. No deep explanation is known. 
 
$\bullet$ The mathematical consistency of the theory, known as ``the
cancellation of the triangle anomalies'', requires that the sum of all electric
charges inside any family is equal to zero. This property has a strong
predictive power.

\subsection{The electromagnetic interactions}

All experimental data are well described by a simple interaction Lagrangian in
which the photon field interacts with a current built out of the fields of
charged particles. 

\be
\label{em1}
{\cal L}_i~\sim ~eA_{\mu}(x)j^{\mu}(x)
\ee

For the spinor matter fields of the Table the current takes the simple form:

\be
\label{em2}
j^{\mu}(x) = \sum_i q_i \bar{\Psi}_i(x)\gamma^{\mu}\Psi_i(x)
\ee
where $q_i$ is the charge of the field $\Psi_i$ in units of $e$. 

This simple Lagrangian has some remarkable properties, all of which are
verified by experiment:

$\bullet$ $j$ is a vector current. The interaction conserves separately $P$, $C$ and
$T$. 

$\bullet$ The current is diagonal in flavour space. 

$\bullet$ More complex terms, such as $j^{\mu}(x)j_{\mu}(x),~~~~\partial A(x)
  \bar{\Psi}(x)... \Psi(x), ....$ are absent, although they do not seem to be
  forbidden by any known property of the theory. All these terms, as well as
  all others we can write, share one common property: In a four-dimensional
  space-time, their canonical dimension is larger than four. We can easily
  show that the resulting quantum field theory is {\it
    non-renormalisable}. For some reason, Nature does not like
  non-renormalisable theories. 

Quantum electrodynamics, the quantum field theory described by the Lagrangian
(\ref{em1}) and supplemented with the programme of renormalisation, is one of
the most successful physical theories. Its agreement with experiment is
spectacular. For years it was the prototype for all other theories. The
Standard Model is the result of the efforts to extend the ideas and methods of
the electromagnetic interactions to all other forces in physics.

\subsection{The weak interactions}

They are mediated by massive vector bosons. When the Standard Model was proposed their very existence, as well as their number, was unknown. But today we know that there exist three, two which are electrically charged and one neutral: $W^+$, $W^-$ and $Z^0$. Like the photon, their couplings to matter are described by current operators:

\be
\label{wi1}
{\cal L}_i~\sim
~V_{\mu}(x)j^{\mu}(x)~~;~~V_{\mu}~:~W_{\mu}^+,~~W_{\mu}^-,~~Z_{\mu}^0
\ee
where the weak currents are again bi-linear in the
    fermion fields: $\bar{\Psi}...\Psi$. Depending on the corresponding vector
    boson, we distinguish two types of weak currents: {\it the charged
      current}, coupled to $W^+$ and $W^-$ and the {\it neutral current}
    coupled to $Z^0$. They have different properties:

{\it The charged current:} 

$\bullet$ Contains only left-handed fermion fields:

\be
\label{wi2}
j_{\mu}\sim \bar{\Psi}_L\gamma_{\mu}\Psi_L \sim
         \bar{\Psi}\gamma_{\mu}(1+\gamma_5)\Psi
\ee

$\bullet$ It is non-diagonal in the quark flavour space.

$\bullet$ The coupling constants are complex. 

{\it The neutral current:}

$\bullet$ Contains both left- and right-handed fermion fields: 

\be
\label{wi3}
j_{\mu}\sim
         C_L\bar{\Psi}_L\gamma_{\mu}\Psi_L+C_R\bar{\Psi}_R\gamma_{\mu}\Psi_R
\ee

$\bullet$ It is diagonal in the quark flavour space.

With these currents weak interactions have some properties which differ from those of the electromagnetic ones:

$\bullet$ Weak interactions violate $P$, $C$ and $T$

$\bullet$ Contrary to the photon, the weak vector bosons are self-coupled. The
nature of these couplings is predicted theoretically in the framework of gauge
theories and it has been determined experimentally.

$\bullet$ A new element has been added recently to the experimental landscape: We have good evidence for the existence of a new particle, compatible with what theorists have called {\it the Higgs boson}, although its properties have not yet been studied in detail. 

It is this kind of interactions that the Standard Model is supposed to
describe.

\section{Gauge symmetries}

\subsection{The concept of symmetry}

In Physics the concept of a Symmetry follows from the assumption that a certain quantity is
not measurable. As a result the equations of motion should not depend on this quantity. We know from the general properties of Classical Mechanics that this implies the existence of conserved quantities. This relation between symmetries and conservation laws, epitomised by Noether's theorem, has been one of the most powerful tools in deciphering the properties of physical theories.

Some simple examples are given by the symmetries of space and time. The assumption that the position of the origin of the coordinate system is not physically measurable implies the invariance of the equations under space translations and the conservation of momentum. In the same way we obtain the conservation laws of energy (time translations) and angular momentum (rotations). We can also distinguish between symmetries under {\it continuous transformations}, such as translations and rotations, and {\it discrete} ones, such as space, or time, inversions. Noether's theorem applies to the first. 

All these symmetries of space and time are {\it geometrical} in the common sense of the word, easy to understand and visualise. During the last century we were led to consider two abstractions, each one of which has had a profound influence in our way of thinking the fundamental interactions. Reversing the chronological order, we shall introduce first the idea of {\it internal} symmetries and, second, that of local or {\it gauge} symmetries. 

\subsection{Internal symmetries}

We call internal symmetries those whose transformation parameters do not affect the point of space and time $x$. The concept of such symmetries can be presented already in classical physics, but it becomes natural in quantum mechanics and quantum field theory. The simplest example is the phase of the wave function. We know that it is not a measurable quantity, so the theory must be invariant under a change of phase. This is true for both relativistic or non-relativistic quantum mechanics. The equations of motion (Dirac or Schr\"odinger), as well as the normalisation condition, are invariant under the transformation:

\be
\label{sym1}
\Psi(x) \rightarrow e^{i\theta}\Psi(x)
\ee

The transformation leaves the space-time point invariant, so it is an internal symmetry. Through Noether's theorem, invariance under (\ref{sym1}) implies the conservation of the probability current. 

The phase transformation (\ref{sym1}) corresponds to the Abelian group $U(1)$. In 1932 Werner Heisenberg enlarged the concept
to a non-Abelian symmetry with the introduction of isospin. The assumption is that strong interactions are invariant under a  group of $SU(2)$ transformations in which the proton and the neutron form a doublet $N(x)$:

\be
\label{sym2}
N(x)= \left (\begin{array}{c}p(x)\\n(x)\end{array}\right ) ~~~;~~~ N(x) \rightarrow e^{i\vec{\tau}\cdot \vec{\theta}}N(x)
\ee
where $\vec{\tau}$ are proportional to the Pauli matrices and $\vec{\theta}$ are the three angles of a general rotation in a three dimensional Euclidean space. Again, the transformations do not apply on the points of ordinary space. 

Heisenberg's iso-space is three dimensional, isomorphic to our physical space. With the discovery of new internal symmetries the idea was generalised to multi-dimensional internal spaces. The space of Physics, {\it i.e.} the space in which all symmetry transformations apply, became an abstract mathematical concept with non-trivial geometrical and topological properties. Only a part of it, the three-dimensional Euclidean space, is directly accessible to our senses.

\subsection{Gauge symmetries}

\begin{figure}
\centering
\epsfxsize=6cm
\epsffile{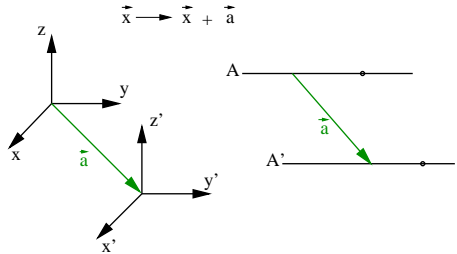}
\caption {A space translation by a constant vector $\vec{a}$} \label{gltr}
\end{figure}

The concept of a local, or gauge, symmetry was introduced by Albert Einstein in his quest for the theory of General Relativity\footnote{It is also present in classical electrodynamics if one considers the invariance under the change of the vector potential $A_\mu (x) \rightarrow A_\mu (x) -  {\partial}_{\mu} \theta (x) $ with $\theta$ an arbitrary function, but before the introduction of quantum mechanics, this aspect of the symmetry was not emphasised.}. Let us come back to the example of space translations, as shown in Figure \ref{gltr}.

\begin{figure}
\centering
\epsfxsize=6cm
\epsffile{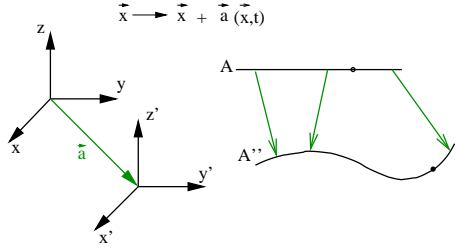}
\caption {A space translation by a vector $\vec{a}(x)$} \label{loctr}
\end{figure}

The figure shows that, if A is the trajectory of a free particle in the (x,y,z) system, its image,
A', is also a possible trajectory of a free particle in the new system. The dynamics of free particles is invariant under space translations by a constant vector. It is a {\it {global}}
invariance, in the sense that the parameter $\vec{a}$ is
independent of the space-time point $x$. Is it possible to extend this
invariance to a {\it {local}} one, namely one in which $\vec{a}$ is replaced by
an arbitrary function of $x$; $\vec{a}(x)$? One calls usually the transformations in which the parameters are functions of the space-time point $x$ {\it gauge transformations}\footnote{This strange terminology is due to Hermann Weyl. In 1918 he attempted to enlarge diffeomorphisms to local scale transformations and he called them, correctly, {\it gauge transformations}. The attempt was unsuccessful, but, when in 1929 he developed the theory for the Dirac electron, although the theory is no more scale
invariant, he still used the term  gauge invariance, a term which has
survived ever since. }. There may be various, essentially
aesthetic, reasons for which one may wish to extend a global invariance to a gauge one. In physical terms, one
may argue that the formalism should allow for a local definition of the origin of the coordinate system,
since the latter is an unobservable quantity. From the mathematical point of view local transformations produce a much richer and more interesting structure. Whichever
one's motivations may be, physical or mathematical, it is clear that
the free particle dynamics is not invariant under translations in which $\vec{a}$ is replaced 
by $\vec{a}(x)$. This is shown schematically in Figure \ref{loctr}. 

We see that no free particle, in its right minds, would follow the trajectory A''. This means that, for A'' to be a trajectory, the particle must be subject to external forces. Can we determine these forces? The question sounds purely geometrical without any obvious physical meaning, so we expect a mathematical answer with no interest for Physics. The great surprise is that the resulting theory which is invariant under local translations turns out to be Classical General
Relativity, one of the four fundamental forces in Nature. Gravitational interactions have such a geometric origin. In fact,
the mathematical formulation of Einstein's original motivation to extend the
Principle of Equivalence to accelerated frames, is precisely the requirement
of local invariance. Historically, many mathematical techniques which are used
in today's gauge theories were developed in the framework of General
Relativity.

\vskip0.5cm

The gravitational forces are not the only ones which have a geometrical
  origin. Let us come back to the example of the quantum mechanical phase. It is clear that neither the Dirac nor the Schr\"odinger equation are invariant under a local change of phase $\theta (x)$. To be precise, let us consider the free Dirac Lagrangian:

\be
\label{sym3}
\mathcal{L} = \bar{\Psi}(x) (i \partial \!\!\!/ - m ) \Psi(x)
\end{equation}

It is not invariant under the transformation:

\be
\label{phasetr}
\Psi(x) \rightarrow e^{i\theta (x)}\Psi(x)
\ee

The reason is the presence of the derivative term in
(\ref{sym3}) which gives rise to a term proportional to
${\partial}_{\mu}\theta (x)$. In order to restore invariance, one must modify
(\ref{sym3}), in which case it will no longer describe a free Dirac field;
invariance under gauge transformations leads to the introduction of
interactions. Both physicists and mathematicians know the answer to the
particular case of (\ref{sym3}): one introduces a new field $A_\mu (x)$ and
replaces the derivative operator ${\partial}_{\mu}$ by a ``covariant
derivative'' $D_{\mu}$  given by: 

\begin{equation}
\label{covder}
D_{\mu} = {\partial}_{\mu} + i e A_\mu
\end{equation}

\noindent where $e$ is an arbitrary real constant. $D_{\mu}$ is called ``covariant'' because it satisfies

\begin{equation}
\label{covd}
D_{\mu} [e^{i \theta (x)} \Psi (x)] = e^{i \theta (x)} D_{\mu} \Psi (x)
\end{equation}

\noindent valid if, at the same time, $A_\mu (x)$ undergoes the transformation:

\begin{equation}
\label{abelgauge}        
A_\mu (x) \rightarrow A_\mu (x) - {\frac {1}{e}} {\partial}_{\mu} \theta (x)   
\end{equation}

The Dirac Lagrangian density becomes now:

\begin{equation}
\label{LDiracinv}
\mathcal{L} = \bar{\Psi}(x) (i D \!\!\!/ - m ) \Psi(x) =  \bar{\Psi}(x) (i \partial \!\!\!/ -eA \!\!\!/ - m ) \Psi(x) 
\end{equation}

It is invariant under the gauge transformations (\ref{phasetr}) and (\ref{abelgauge}) and describes the interaction of a charged spinor field with an external electromagnetic field! Replacing the derivative operator by the covariant derivative
turns the Dirac equation into
the same equation in the presence of an external electromagnetic field. Electromagnetic interactions admit the same geometrical interpretation\footnote{The same applies to the Schr\"odinger equation. In fact, this was done first by V. Fock in 1926, immediately after Schr\"odinger's original publication.}. We can complete the picture by including the degrees of freedom of the electromagnetic field itself and add to (\ref{LDiracinv}) the corresponding Lagrangian density. Again, gauge invariance determines its form uniquely and we are led to the well-known result:

\begin{equation}
\label{LDiracinv1}
\mathcal{L} = -\frac{1}{4}F_{\mu \nu}(x)F^{\mu \nu}(x) +\bar{\Psi}(x) (i D \!\!\!/ - m ) \Psi(x) 
\ee
with
\be
\label{Fmunu}
F_{\mu \nu}(x)=\partial_{\mu}A_{\nu}(x)-\partial_{\nu}A_{\mu}(x)
\ee

The constant $e$ we introduced is the electric charge, the coupling strength
of the field $\Psi$ with the electromagnetic field. Notice that a second field
${\Psi}'$ will be coupled with its own charge $e'$. 

Let us summarise: We started with a  theory invariant under a group
$U(1)$ of global phase transformations. The extension to a local invariance can
be interpreted as a $U(1)$ symmetry at each point $x$. In a qualitative way we
can say that gauge invariance induces an invariance under $U(1)^{\infty}$. We
saw that this extension, a purely geometrical requirement, implies the
introduction of new interactions. The surprising result here is that these
``geometrical'' interactions describe the well-known electromagnetic forces.
\vskip 0.5cm

The extension of the formalism of gauge theories to non-Abelian groups is not
trivial and was first discovered by trial and error. Here we shall restrict
ourselves to internal symmetries which are simpler to analyse and they are the
ones we 
shall apply to particle physics outside gravitation.  

Let us consider a classical field theory given by a Lagrangian density $\cal
L$. It depends on a set of $N$ fields $\psi^i (x)$, $i = 1,...,r$ and their
first derivatives. The Lorentz transformation properties of these fields will
play no role in this discussion. We assume that the $\psi$'s transform linearly
according to an $r$-dimensional  representation, not necessarily irreducible,
of a compact, simple,  Lie group $G$ which does not act on the space-time
point $x$. 

\begin{equation}
\label{gentrans} 
\Psi = \left (\begin{array}{c}\psi^1 \\ \vdots \\ \psi^r \end{array} \right )
\hskip 2cm  \Psi (x) \rightarrow U(\omega)\Psi (x) \hskip 2cm \omega \in G
\end{equation}

\noindent where $U(\omega)$ is the matrix of the representation of $G$. In
fact, in these lectures we shall be dealing only with perturbation theory and
it will be sufficient to look at transformations close to
the identity in $G$.

\begin{equation}
\label{gentransf} 
\Psi (x) \rightarrow e^{i\Theta}\Psi(x) \hskip 2cm \Theta = \sum_{a = 1}^{m}
\theta ^{a} T^{a}  
\end{equation}
where the $\theta ^{a}$'s are a set of $m$  constant parameters, and
the $T^{a}$'s are $m$ $r\times r$ matrices representing the $m$ generators of
the Lie algebra of $G$. They satisfy the commutation rules: 

\begin{equation}
\label{commut}
\left [T^{a} , T^{b} \right ] = i f^{abc} T^c 
\end{equation}

The $f$'s are the structure constants of $G$ and a summation over
repeated indices is understood. The normalisation of the structure constants
is usually fixed by requiring that, in the fundamental representation, the
corresponding matrices of the generators $t^a$ are normalised such as: 

\begin{equation}
\label{norm}
Tr \left (t^{a} t^{b} \right) = \frac {1}{2} \delta ^{ab}
\end{equation}

The Lagrangian density ${\cal L} (\Psi , \partial \Psi )$ is assumed to be
invariant under the global transformations (\ref{gentransf}) or
(\ref{gentrans}). As was done for the Abelian case, we wish to find a new
$\cal L$, invariant under the corresponding gauge transformations in which the
$\theta ^{a}$'s of (\ref{gentransf}) are arbitrary functions of $x$. In the
same qualitative sense, we look for a theory invariant under
$G^{\infty}$. This problem, stated the way we present it here, was first solved by trial and error for the case of
$SU(2)$ by C.N. Yang and R.L. Mills in 1954. They  gave 
the underlying physical motivation and these theories are called since
``Yang-Mills theories''. The steps are direct generalisations of the ones
followed in the Abelian case. We need a gauge field, the analogue of the
electromagnetic field, to transport the information contained in
(\ref{gentransf}) from point to point. Since we can perform $m$ independent
transformations, the number of generators in the Lie algebra of $G$, we need
$m$ gauge fields $A_{\mu}^{a}(x)$, $a=1,...,m$. It is easy to show that they
belong to the adjoint representation of $G$. Using the matrix representation
of the generators we can cast $A_{\mu}^{a}(x)$ into an $r \times r$ matrix: 

\begin{equation}
\label{gaugefield}
{\cal A}_{\mu}(x) = \sum_{a = 1}^{m} A_{\mu}^{a}(x) T^{a} 
\end{equation}

The covariant derivatives can now be constructed as: 

\begin{equation}
\label{gencovder}      
{\cal D}_{\mu} = {\partial}_{\mu} + i g{\cal A}_\mu
\end{equation}

\noindent with $g$ an arbitrary real constant. They satisfy:

\begin{equation}
\label{gencovderiv}
{\cal D}_{\mu} e^{i\Theta (x)}\Psi (x) = e^{i\Theta (x)}{\cal D}_{\mu}\Psi (x)
\end{equation}

\noindent provided the gauge fields transform as:

\begin{equation}
\label{gaugtrans}
{\cal A}_{\mu}(x) \rightarrow e^{i\Theta (x)}{\cal A}_{\mu}(x)e^{-i\Theta (x)}
+ \frac {i}{g} \left (\partial_{\mu}e^{i\Theta (x)}\right )e^{-i\Theta (x)}  
\end{equation}

The Lagrangian density ${\cal L}(\Psi ,{\cal D}\Psi)$ is invariant under the
gauge transformations (\ref{gentransf}) and (\ref{gaugtrans}) with an
$x$-dependent $\Theta$, if ${\cal L}(\Psi ,{\partial}\Psi)$ is invariant under
the corresponding global ones (\ref{gentrans}) or (\ref{gentransf}). As was
done with the electromagnetic field, we can include the degrees of freedom of
the new gauge fields by adding to the Lagrangian density a gauge invariant
kinetic term. It turns out that it is slightly more complicated than $F_{\mu
  \nu}$ of the Abelian case. Yang and Mills computed it for $SU(2)$ but, in
fact, it is uniquely determined by geometry plus some obvious requirements,
such as absence of higher order derivatives. The result is given by:

\begin{equation}
\label{curv}
{\cal G}_{\mu \nu} = \partial_{\mu} {\cal A}_{\nu} - \partial_{\nu} {\cal A}_{\mu} 
- ig \left [ {\cal A}_{\mu} , {\cal A}_{\nu} \right ]  
\end{equation}

The full gauge-invariant Lagrangian can now be written as:

\begin{equation}
\label{inv}
{\cal L}_{inv} = -\frac {1}{2} Tr {\cal G}_{\mu \nu} {\cal G}^{\mu \nu} + {\cal L}(\Psi ,{\cal D}\Psi)
\end{equation} 

By convention, in (\ref{curv}) the matrix ${\cal A}$ is taken to be:

\begin{equation}
\label{fund}
{\cal A}_{\mu} = A_{\mu}^a t^a
\end{equation}

\noindent where we recall that the $t^a$'s are the matrices representing the
generators in the fundamental representation. It is only with this convention
that the kinetic term in (\ref{inv}) is correctly normalised. In terms of the
component fields $A_{\mu}^a$, ${\cal G}_{\mu \nu}$ reads: 

\begin{equation}
\label{comp}
{\cal G}_{\mu \nu} = G_{\mu \nu}^a t^a \hskip 2cm G_{\mu \nu}^a = \partial_{\mu}A_{\nu}^a - \partial_{\nu}A_{\mu}^a + g f^{abc} A_{\mu}^b A_{\nu}^c
\end{equation}

Under a gauge transformation ${\cal G}_{\mu \nu}$ transforms like a member of the adjoint representation.

\begin{equation}
\label{adj}
{\cal G}_{\mu \nu} (x) \rightarrow e^{i{\theta}^{a}(x)t^a}\hskip0.2cm {\cal
  G}_{\mu \nu} (x)\hskip0.2cm e^{-i{\theta}^{a}(x)t^a} 
\end{equation}

This completes the construction of the gauge invariant Lagrangian.
We add some remarks: 

$\bullet$ As it was the case with the electromagnetic field, the Lagrangian 
(\ref{inv}) does not contain terms proportional to $A_{\mu}A^{\mu}$. This
means that, under the usual quantisation rules, the gauge
fields describe massless particles. 

$\bullet$ Since ${\cal G}_{\mu \nu}$ is not linear in the fields ${\cal A}_{\mu}$,
the ${\cal G}^2$ term in (\ref{inv}), besides the usual kinetic term which is
bilinear in the fields, contains tri-linear and quadri-linear terms. In
perturbation theory they will be treated as coupling terms whose strength is
given by the coupling constant $g$. In other words, the non-Abelian gauge
fields are self-coupled while the Abelian (photon) field is not. A Yang-Mills
theory, containing only gauge fields, is still a dynamically rich quantum
field theory while a theory with the electromagnetic field alone is a trivial
free theory. 

$\bullet$ The same coupling constant $g$ appears in the covariant derivative of
the fields $\Psi$ in (\ref{gencovder}). This simple consequence of gauge
invariance has an important physical application: if we add another field $
{\Psi}'$, its coupling strength with the gauge fields will still be given by
the same constant $g$. Contrary to the Abelian case studied before, if
electromagnetism is part of a non-Abelian simple group, gauge
invariance implies charge quantisation. 

$\bullet$ The above analysis can be extended in a straightforward way to the case
where the group $G$ is the product of simple groups $G = G_1 \times ...\times
G_n$. The only difference is that one should introduce $n$ coupling constants
$g_1,...,g_n$, one for each simple factor. Charge quantisation is still true
inside each subgroup, but charges belonging to different factors are no more
related. 

$\bullet$ The situation changes if one considers non semi-simple groups, where one,
or more, of the factors $G_i$ is Abelian. In this case the associated coupling
constants can be chosen different for each field and the corresponding Abelian
charges are not quantised.

\vskip 0.5cm

As we alluded to above, gauge theories have a deep geometrical meaning. In
order to get a better understanding of this property without entering into
complicated issues of differential geometry, it is instructive to consider a
reformulation of the theory replacing the continuum of space-time with a four
dimensional Euclidean lattice. We can do that very easily. Let us consider, for simplicity, a lattice with hypercubic symmetry. The space-time point $x_{\mu}$ is replaced by:

\be
\label{latt1}
x_{\mu} \rightarrow n_{\mu}a
\ee
where $a$ is a constant length, (the lattice spacing), and $n_{\mu}$ is a
$d$-dimensional vector with components $n_{\mu}= (n_1, n_2, ...,n_d)$
which take integer values $0\leq n_{\mu} \leq N_{\mu}$. $N_{\mu}$ is the
number of points of our lattice in the direction $\mu$. The total number of
points, {\it i.e.} the volume of the system, is given by $V\sim \prod_{\mu
  =1}^dN_{\mu}$. The presence of $a$ introduces an ultraviolet, or short
distance, cut-off because all momenta are bounded from above by $2\pi/a$. The
presence of $N_{\mu}$ introduces an infrared, or large distance cut-off
because the momenta are also bounded from below by $2\pi/Na$, where $N$ is the
maximum of $N_{\mu}$.  The infinite volume continuum  space is recovered at
the double limit $a\rightarrow 0$ and $N_{\mu} \rightarrow \infty$.

The dictionary between quantities defined in the continuum and the corresponding ones on the lattice is easy to establish (we take the lattice spacing $a$ equal to one):

$\bullet$ A field $\Psi(x) ~~~\Rightarrow ~~~\Psi_n$

\noindent where the field $\Psi$ is an $r$-component column vector as in
equation (\ref{gentrans}).

$\bullet$ A local term such as $\bar{\Psi}(x)\Psi(x)~~~\Rightarrow ~~~\bar{\Psi}_n\Psi_n$

$\bullet$ A derivative $\partial_{\mu} \Psi(x) ~~~\Rightarrow ~~~(\Psi_n-\Psi_{n+\mu})$

\noindent where $n+\mu$ should be understood as a unit vector joining the point $n$ with its nearest neighbour in the direction $\mu$. 

$\bullet$ The kinetic energy term\footnote{We write here the expression for spinor fields which contain only first order derivatives in the kinetic energy. The extension to scalar fields with second order derivatives is obvious.} $\bar{\Psi}(x)\partial_{\mu} \Psi(x) ~~~\Rightarrow ~~~\bar{\Psi}_n\Psi_n-\bar{\Psi}_n\Psi_{n+\mu}$

We may be tempted to write similar expressions for the gauge fields, but we
must be careful with the way gauge transformations act on the lattice. Let us
repeat the steps we followed in the continuum: Under gauge transformations a
field transforms as:

$\bullet$ Gauge transformations $\Psi (x) \rightarrow e^{i\Theta
  (x)}\Psi(x)~~~\Rightarrow ~~~\Psi_n \rightarrow e^{i\Theta_n}\Psi_n$

All local terms of the form $\bar{\Psi}_n\Psi_n$ remain invariant but the part
of the kinetic energy which couples fields at neighbouring points does not.

$\bullet$ The kinetic energy $\bar{\Psi}_n\Psi_{n+\mu} \rightarrow
\bar{\Psi}_n e^{-i\Theta_n}e^{i\Theta_{n+\mu}}\Psi_{n+\mu}$

\noindent which shows that we recover the problem we had with the derivative
operator in the continuum. In order to restore invariance we must introduce a
new field, which is an $r$-by-$r$ matrix, and which has indices $n$ and
$n+\mu$. We denote it by $U_{n,n+\mu}$ and we shall impose on it the constraint $U_{n,n+\mu}=U^{-1}_{n+\mu,n}$.  Under a
gauge transformations, $U$ transforms as:

\be
\label{gtr}
U_{n,n+\mu}~~\rightarrow~~e^{i\Theta_n}U_{n,n+\mu}e^{-i\Theta_{n+\mu}}
\ee

With the help of this gauge field we write the kinetic energy term with the
covariant derivative on the lattice as:

\be
\label{gtr1}
\bar{\Psi}_n~U_{n,n+\mu}~\Psi_{n+\mu}
\ee
which is invariant under gauge transformations. 

$U$ is an element of the gauge group but we can show that, at the continuum
limit and for an infinitesimal transformation, it reproduces correctly $A_{\mu}$, which belongs to the Lie algebra of
the group. Notice that, contrary to the field $\Psi$,  $U$ does not live on a
single lattice point, but it has two indices, $n$ and $n+\mu$, in other words
it lives on the oriented link joining the two neighbouring points. We see here
that the mathematicians are right when they do not call the gauge field ``a
field'' but ``a connection''. 

In order to finish the story we want to obtain an expression for the kinetic
energy of the gauge field, the analogue of $Tr {\cal
  G}_{\mu \nu}(x) {\cal G}^{\mu \nu}(x)$,  on the lattice. As for the continuum, the guiding principle is gauge invariance. Let us
consider two points on the lattice $n$ and $m$. We shall call a path $p_{n,m}$
on the lattice a sequence of oriented links which join continuously the two
points. Consider next the product of the gauge fields $U$ along all the
links of the path $p_{n,m}$:

\be
\label{latg11}
P^{(p)}(n,m)=\prod_pU_{n,n+\mu}...U_{m-\nu,m}
\ee

Using the transformation rule (\ref{gtr}), we see that $P^{(p)}(n,m)$
transforms as:

\be
\label{latg12}
P^{(p)}(n,m)\rightarrow e^{i\Theta_n}P^{(p)}(n,m)e^{-i\Theta_m}
\ee

It follows that if we consider a closed path $c=p_{n,n}$ the quantity Tr$P^{(c)}$
is gauge invariant. The simplest closed path for a hypercubic lattice has four
links and it is called {\it plaquette.} The correct form of the Yang-Mills
action on the lattice can be written in terms of the sum of Tr$P^{(c)}$ over
all plaquettes.

\section{Spontaneous symmetry breaking}

Since gauge theories appear to predict the existence of massless gauge bosons,
when they were first proposed they did not seem to have any direct application
to particle physics outside 
electromagnetism. It is this handicap which plagued gauge theories for many
years. In this section we shall present a seemingly unrelated phenomenon
which, however, will turn out to provide the answer. 

\vskip 0.5cm

An infinite system may exhibit the phenomenon of phase transitions. It often
implies a reduction in the symmetry of the ground state. A field theory is a
system with an infinite number of degrees of freedom, so, it is not surprising
that field theories may also show the phenomenon of phase transitions. Indeed, in
many cases, we encounter at least 
two phases: 

$\bullet$ {\it The unbroken}, or, {\it the Wigner phase}: The symmetry is manifest
in the spectrum of the theory whose excitations form irreducible
representations of the symmetry group. For a gauge theory the vector gauge
bosons are massless and belong to the adjoint representation. But we have good
reasons to believe that, for non-Abelian gauge theories, a strange 
phenomenon occurs in this phase:  all physical states are singlets of the
group. All non-singlet states, such as those corresponding to the gauge
fields, are supposed to be {\it confined}, in the sense that they do not
appear as physically realisable asymptotic states.

$\bullet$ {\it The spontaneously broken phase}: Part of the symmetry is
hidden from the spectrum. For a gauge theory, some of the gauge bosons become
massive and appear as physical states.

It is this kind of phase transition that we want to study in this section. 

\subsection{An example from classical mechanics}

A very simple  example is provided by the problem of the bent rod. Let a
cylindrical rod be charged as in Figure \ref{SpBr1}. The problem is obviously symmetric
under rotations around the $z$-axis. Let $z$ measure the distance from the
basis of the rod,
and $X(z)$ and $Y(z)$ give the deviations, along the $x$ and $y$ directions
respectively, of the axis of the rod at the point $z$ from the symmetric
position. For small deflections the equations of elasticity take the form:

\begin{figure}
\centering
\epsfxsize=6cm
\epsffile{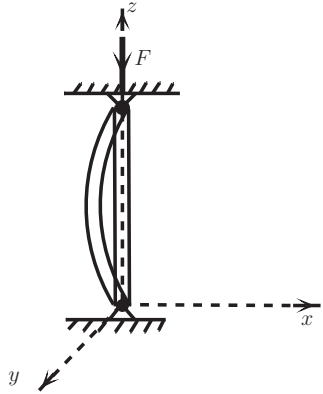}
\caption {A cylindrical rod bent under a force $F$ along its symmetry axis.} \label{SpBr1}
\end{figure}

\begin{equation}
\label{elast1}
IE\frac{d^4 X}{dz^4}~+~F\frac{d^2 X}{dz^2}~=~0~~~~;~~~~IE\frac{d^4 Y}{dz^4}~+~F\frac{d^2 Y}{dz^2}~=~0
\end{equation}
where $I = \pi R^4 /4$ is the moment of inertia of the rod and $E$ is the
Young modulus. It is obvious that the system (\ref{elast1}) always
possesses a symmetric solution $X = Y = 0$. However, we can also
look for asymmetric solutions of the general form:
$X = A + Bz + C$sin$kz + D$cos$kz$ with $k^2 = F/EI$, which satisfy
the boundary conditions $X = X'' = 0$ at $z = 0$ and $z = l$. We find
that such solutions exist,
$X = C$sin$kz$, provided $kl = n\pi$ ; $n$ = 1, ... . The first such
solution appears
when $F$ reaches a critical value $F_{cr}$                   given by:

\begin{equation}
\label{elast2}
F_{cr}=\frac{{\pi}^2 EI}{l^2}
\end{equation}

The appearance of these solutions is already an indication of
instability and, indeed, a careful study of the stability problem
proves that the non-symmetric solutions correspond to lower
energy. From that point Eqs. (\ref{elast1}) are no longer valid, because
they only apply to small deflections, and we must use the general
equations of elasticity. The result is that this instability of
the symmetric solution occurs for all values of $F$ larger than $F_{cr}$

   What has happened to the original symmetry of the equations?
It is still hidden in the sense that we cannot predict in which
direction in the $x-y$ plane the rod is going to bend. They all
correspond to solutions with precisely the same energy. In other
words, if we apply a symmetry transformation (in this case a
rotation around the $z$-axis) to an asymmetric solution, we obtain
another asymmetric solution which is degenerate with the first
one.

We call such a symmetry ``spontaneously broken'', and in this
simple example we see all its characteristics: 

$\bullet$ There exists a
critical point, i.e. a critical value of some external quantity which we can
vary freely, (in this
case the external force $F$; in several physical systems it is the
temperature) which determines whether spontaneous symmetry
breaking will take place or not. Beyond this critical point:

$\bullet$  The symmetric solution becomes unstable.

$\bullet$ The ground state becomes degenerate.
\vskip 0.5 cm

   There exist a great variety of physical systems, both in
classical and quantum physics, exhibiting spontaneous symmetry
breaking, but we will not describe any other one here. The
Heisenberg ferromagnet is a good example to keep in mind, because
we shall often use it as a guide, but no essentially new
phenomenon appears outside the ones we saw already. Therefore, we
shall go directly to some field theory models.

\subsection{A simple field theory model}

Let $\phi(x)$ be a complex scalar field whose dynamics is
described by the Lagrangian density:
                                                    
\begin{equation}
\label{spsymbr1}
{\cal {L}}_1~=~(\partial_{\mu}\phi)(\partial^{\mu}\phi^*)-M^2 \phi \phi^*
-\lambda (\phi \phi^*)^2
\end{equation}
where ${\cal {L}}_1$ is a classical Lagrangian density and $\phi(x)$ is a classical field. No quantisation is considered for the moment. (\ref{spsymbr1})
is invariant under the group $U(1)$ of global
transformations:

\begin{equation}
\label{spsymbr2}
\phi(x)~~\rightarrow ~~e^{i\theta}\phi(x)
\end{equation}

To this invariance corresponds the current $j_{\mu}\sim \phi\partial_{\mu}\phi^*-\phi^*\partial_{\mu}\phi$ whose conservation can be verified using the equations of motion.

We are interested in the classical field configuration which minimises the energy of the system. We thus compute the Hamiltonian density given by
                                         
\begin{equation}
\label{spsymbr3}                                                        
{\cal {H}}_1~=~(\partial_0 \phi)(\partial_0 \phi^*)+(\partial_i
\phi)(\partial_i \phi^*)+V(\phi)
\end{equation}

\begin{equation}
\label{spsymbr4}
V(\phi)=M^2 \phi \phi^*
+\lambda (\phi \phi^*)^2
\end{equation}

   The first two terms of  ${\cal {H}}_1$ are positive definite. They can only vanish
for $\phi$ = constant. Therefore, the ground state of the system
corresponds to $\phi$ = constant = minimum of $V(\phi)$. $V$ has a minimum
only if  $\lambda >$ 0. In this case the position of the minimum depends
on the sign of $M^2$. (Notice that we are still studying a classical
field theory and $M^2$ is just a parameter. One should not be misled by the
notation into thinking that $M$ is a ``mass'' and $M^2$ is necessarily
positive).

\begin{figure}
\centering
\epsfxsize=12cm
\epsffile{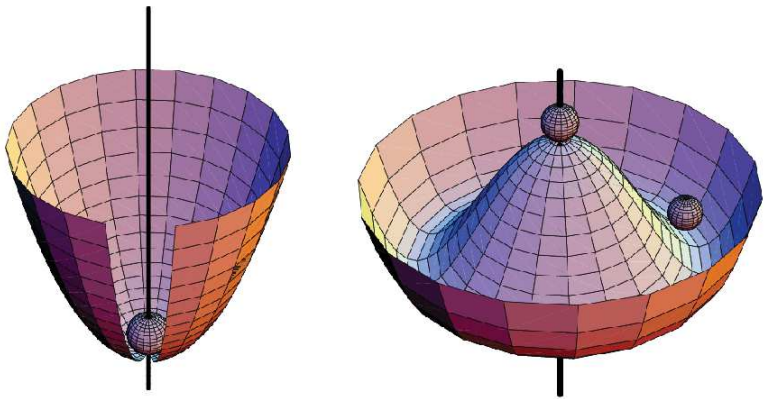}
\caption {The potential $V(\phi)$ with $M^2\geq 0$ (left) and $M^2<$0 (right)} \label{SpBr}
\end{figure}

For $M^2 >$ 0 the minimum is at $\phi$ =  0 (symmetric solution, shown in the
left side of Figure \ref{SpBr}),
but for $M^2 <$ 0 there is a whole circle of minima at the complex
$\phi$-plane with radius $v=(-M^2 /2\lambda)^{1/2}$ (Figure \ref{SpBr}, right side). Any point on the
circle corresponds to a spontaneous breaking of (\ref{spsymbr2}). 

We see that:

$\bullet$ The critical point is $M^2$ = 0;

$\bullet$    For $M^2 >$ 0 the symmetric solution is stable;

$\bullet$    For $M^2 <$ 0 spontaneous symmetry breaking occurs.

 Let us choose $M^2 <$ 0 . In order to reach the stable
solution we translate the field $\phi$. It is clear that there is no
loss of generality by choosing a particular point on the circle,
since they are all obtained from any given one by applying the
transformations (\ref{spsymbr2}). Let us, for convenience, choose the point
on the real axis in the $\phi$-plane. We thus write:

\begin{equation}
\label{spsymbr5}                
\phi(x)=\frac{1}{\sqrt{2}}\left[v+\psi(x)+i\chi(x)\right]
\end{equation}

Bringing (\ref{spsymbr5}) in (\ref{spsymbr1}) we find

\begin{equation}
\begin{split}     
{\cal {L}}_1(\phi)~\rightarrow~{\cal {L}}_2(\psi
,\chi) & =\frac{1}{2}(\partial_{\mu}\psi)^2+\frac{1}{2}(\partial_{\mu}\chi)^2-
\frac{1}{2}(2\lambda v^2)\psi^2 \\
 & -\lambda v\psi(\psi^2+\chi^2)-\frac{\lambda}{4}(\psi^2+\chi^2)^2
\end{split}
\label{spsymbr6}
\end{equation}

Notice that ${\cal {L}}_2$ does not contain any term proportional to $\chi^2$,
which is expected since V is locally flat in the $\chi$ direction. A second
remark concerns the arbitrary parameters of the theory. ${\cal {L}}_1$
contains two such parameters, a mass $M$ and a dimensionless coupling constant
$\lambda$. In ${\cal {L}}_2$ we have again the coupling constant $\lambda$ and
a new mass parameter $v$ which is a function of $M$ and $\lambda$. It is
important to notice that, although ${\cal {L}}_2$ contains also trilinear
terms, their coupling strength is not a new parameter but is proportional to
$v\lambda$.  ${\cal {L}}_2$ is still invariant under the transformations with infinitesimal parameter $\theta$: 

\be
\label{spsymbr21}
\delta \psi = -\theta \chi~~~~~;~~~~~\delta\chi=\theta\psi+\theta v
\ee
to which corresponds a conserved current 

\be
\label{spsymbr22}
j_{\mu}\sim \psi\partial_{\mu}\chi-\chi\partial_{\mu}\psi+v\partial_{\mu}\chi
\ee  

The last term, which is linear in the derivative of $\chi$, is characteristic of the phenomenon of spontaneous symmetry breaking.

It should be emphasised here that ${\cal {L}}_1$ and  ${\cal {L}}_2$ are completely
equivalent Lagrangians. They both describe the dynamics of the
same physical system and a change of variables, such as (\ref{spsymbr5}),
cannot change the physics. However, this equivalence is only true
if we can solve the problem exactly. In this case we
shall find the same solution using either of them. However, we
do not have exact solutions and we intend to apply perturbation
theory, which is an approximation scheme. Then the equivalence
is no longer guaranteed and, in fact, perturbation theory has
much better chances to give sensible results using one language
rather than the other. In particular, if we use ${\cal {L}}_1$ as a quantum 
field theory and we decide to apply perturbation theory taking,
as the unperturbed part, the quadratic terms of ${\cal {L}}_1$, we immediately
see that we shall get nonsense. The spectrum of the unperturbed
Hamiltonian would consist of particles with negative square
mass, and no perturbation corrections, at any finite order,
could change that. This is essentially due to the fact that, in
doing so, we are trying to calculate the quantum fluctuations
around an unstable solution and perturbation theory is just not
designed to do so. On the contrary, we see that the quadratic
part of ${\cal {L}}_2$ gives a reasonable spectrum; thus we hope that
perturbation theory will also give reasonable results. Therefore
we conclude that our physical system, considered now as a
quantum system, consists of two interacting scalar particles,
one with mass $m_{\psi}^2=2\lambda v^2$ and the other with $m_{\chi}$ = 0. We
believe that this is the
spectrum  we would have found also starting from ${\cal {L}}_1$, if we
could solve the dynamics exactly.

 The appearance of a zero-mass particle in the quantum version of the model is an example of a
general theorem due to J. Goldstone: To every generator of a
spontaneously broken symmetry there corresponds a massless
particle, called the Goldstone particle. This theorem is just
the translation, into quantum field theory language, of the
statement about the degeneracy of the ground state. The ground state of a system described by a
quantum field theory is the vacuum state, and you need massless
excitations in the spectrum of states in order to allow for the
degeneracy of the vacuum. 

\subsection{Gauge symmetries}

In this section we want to study the consequences of spontaneous symmetry
breaking in the presence of a gauge symmetry. We shall find a very surprising result. When
combined together the two problems, namely the massless gauge bosons on the
one hand and the
massless Goldstone bosons on the other, will solve each other. It is this miracle that we want to present here. We start with the Abelian case.

We look at the model of the previous section  in which the $U(1)$ symmetry (\ref{spsymbr2}) has been promoted to a local symmetry with $\theta \rightarrow \theta(x)$. As we explained already, this implies the introduction of a massless vector field, which we can call the ``photon'' and the interactions are obtained by
replacing the derivative operator $\partial_{\mu}$ by the covariant derivative
$D_{\mu}$ and adding the photon kinetic energy term:

\begin{equation}
\label{spbrg1}
{\cal {L}}_1=-\frac{1}{4}F_{\mu
  \nu}^2+|(\partial_{\mu}+ieA_{\mu})\phi|^2-M^2\phi \phi^*-\lambda(\phi \phi^*)^2 
\end{equation}

${\cal {L}}_1$ is invariant under the gauge transformation:

\begin{equation}
\label{spbrg2}
\phi(x)~~\rightarrow ~~e^{i\theta (x)}\phi(x)~~~~;~~~~A_{\mu}~~\rightarrow ~~A_{\mu}-\frac{1}{e}\partial_{\mu}\theta (x)
\end{equation} 

The same analysis as before shows that for $\lambda > 0$ and $M^2 < 0$
there is a spontaneous breaking of the $U(1)$ symmetry. Replacing
(\ref{spsymbr5}) into (\ref{spbrg1}) we obtain:

\begin{equation}
\begin{split}
{\cal {L}}_1~\rightarrow~{\cal {L}}_2 & =-\frac{1}{4}F_{\mu \nu}^2+\frac{e^2
  v^2}{2}A_{\mu}^2 +evA_{\mu}\partial^{\mu}\chi\\
& +\frac{1}{2}(\partial_{\mu}\psi)^2+\frac{1}{2}(\partial_{\mu}\chi)^2-
\frac{1}{2}(2\lambda v^2)\psi^2 ~~+...
\end{split}
\label{spbrg3}
\end{equation}
where the dots stand for coupling terms which are at least trilinear in the
fields. 

The surprising term is the second one which is proportional to
$A_{\mu}^2$. It looks as though the photon has become massive. Notice
that (\ref{spbrg3}) is still gauge invariant since it is equivalent to
(\ref{spbrg1}). The gauge transformation is now obtained by replacing
(\ref{spsymbr5}) into (\ref{spbrg2}):

\begin{equation}
\begin{split}
& \psi(x)~~\rightarrow ~~{\mathrm {cos}}\theta (x)[\psi(x)+v]-{\mathrm {sin}}\theta (x)\chi(x)-v\\
& \chi(x)~~\rightarrow ~~{\mathrm {cos}}\theta (x) \chi(x)+{\mathrm {sin}}\theta (x)[\psi(x)+v]\\
& A_{\mu}~~\rightarrow ~~A_{\mu}-\frac{1}{e}\partial_{\mu}\theta (x)
\end{split}
\label{spbrg4}
\end{equation}

This means that our previous conclusion, that gauge invariance
forbids the presence of an $A_{\mu}^2$ term, was simply wrong. Such a
term can be present, only the gauge transformation is slightly
more complicated; it must be accompanied by a translation of the
field.

The Lagrangian (\ref{spbrg3}), if taken as a quantum field theory,
seems to describe the interaction of a massive vector particle
($A_{\mu}$) and two scalars, one massive ($\psi$)  and one massless ($\chi$).
However, we can see immediately that something is wrong with
this counting. A warning is already contained in the
 non-diagonal term between $A_{\mu}$ and $\partial^{\mu}\chi$. Indeed, the
 perturbative particle spectrum can be read from the Lagrangian only after we
 have diagonalised the quadratic part. A more direct way to see the trouble is
 to count the apparent
degrees of freedom\footnote{The terminology here is misleading. As we pointed
  out earlier, any field theory, considered as a dynamical system, is a system
  with an infinite number of degrees of freedom. For example, the quantum
  theory of a free neutral scalar field is described by an infinite number of
  harmonic oscillators, one for every value of the three-dimentional
  momentum. Here we use the same term ``degrees of freedom'' to denote the
  independent one-particle states. We know that for a massive spin-$s$
  particle we have $2s+1$ one-particle states and for a massless particle with
  spin different from zero we have only two. In fact, it would have been more
  appropriate to talk about a ($2s+1$)-{\it infinity} and 2-{\it infinity}
  degrees of freedom, respectively.} before and after the translation:

\vskip 0.3 cm
$\bullet$ Lagrangian (\ref{spbrg1}):

\noindent (i) One massless vector field: 2 degrees

\noindent (ii) One complex scalar field: 2 degrees

\noindent Total: 4 degrees

\vskip 0.3 cm
$\bullet$ Lagrangian (\ref{spbrg3}):

\noindent (i) One massive vector field: 3 degrees

\noindent (ii) Two real scalar fields: 2 degrees

\noindent Total: 5 degrees  
\vskip 0.3 cm

Since physical degrees of freedom cannot be created by a simple
change of variables, we conclude that the Lagrangian (\ref{spbrg3}) must
contain fields which do not create physical particles.
This is indeed the case, and we can exhibit a transformation which makes
 the unphysical fields disappear. Instead of parametrising the complex field
 $\phi$ by its real and imaginary parts, let us choose its modulus and its
 phase. The choice is dictated by the fact that it is a change of phase that
 describes the motion along the circle of the minima of the potential $V(\phi)$. We
 thus write:

\begin{equation}
\label{spbrg5}
\phi(x)=\frac{1}{\sqrt{2}}[v+\rho(x)]e^{i\zeta(x)/v}~~~~;~~~~A_{\mu}(x)=B_{\mu}(x)-\frac{1}{ev}\partial_{\mu}\zeta(x)
\end{equation}

In this notation, the gauge transformation (\ref{spbrg2}) or (\ref{spbrg4}) is
simply a translation of the field $\zeta$: $\zeta(x) \rightarrow \zeta(x) + v
\theta(x)$. Replacing (\ref{spbrg5}) into (\ref{spbrg1}) we obtain:

\begin{equation}
\begin{split}     
{\cal {L}}_1~\rightarrow~{\cal {L}}_3 &=-\frac{1}{4}B_{\mu \nu}^2+\frac{e^2
  v^2}{2}B_{\mu}^2+\frac{1}{2}(\partial_{\mu}\rho)^2-\frac{1}{2}(2\lambda
  v^2)\rho^2\\
 &-\frac{\lambda}{4}\rho^4+\frac{1}{2}e^2B_{\mu}^2(2v\rho+\rho^2)\\
B_{\mu \nu}&=\partial_{\mu}B_{\nu}-\partial_{\nu}B_{\mu}
\end{split}
\label{spbrg6}
\end{equation}

The field $\zeta(x)$ has disappeared. Formula (\ref{spbrg6}) describes two
massive particles, a vector ($B_{\mu}$) and a scalar ($\rho$). It exhibits no gauge
invariance, since the original symmetry $\zeta(x) \rightarrow \zeta(x) + v
\theta(x)$ is now trivial. 

We see that
we obtained three different Lagrangians describing the same
physical system. ${\cal {L}}_1$ is invariant under the usual gauge
transformation, but it contains a negative square mass and, therefore, it is
unsuitable for quantisation. ${\cal {L}}_2$ is still gauge invariant, but the
transformation law (\ref{spbrg4}) is more complicated. It can be quantised in
a space containing unphysical degrees of freedom. This, by itself, is not a
great obstacle and it occurs frequently. For example, ordinary quantum
electrodynamics is usually quantised in a space involving unphysical
(longitudinal and scalar) photons. In fact, it is ${\cal {L}}_2$, in a
suitable gauge, which is used for general proofs of renormalisability as well
as for practical calculations. Finally ${\cal {L}}_3$ is no longer invariant
under any kind of gauge transformation, but it exhibits clearly the particle
spectrum of the theory. It contains only physical particles and they are all
massive. This is the miracle that was announced earlier. Although we start from a gauge
theory, the final spectrum contains massive particles only. Actually, ${\cal {L}}_3$ can be obtained from ${\cal {L}}_2$ by an
appropriate choice of gauge.

The conclusion so far can  be stated as follows:

In a spontaneously broken gauge theory the gauge vector bosons acquire a mass
and the would-be massless Goldstone bosons decouple and disappear. Their
degrees of freedom are used in order to make possible the transition from
massless to massive vector bosons.

\vskip 0.5cm

The extension to the non-Abelian case is straightforward. Let us consider a
gauge group $G$ with $m$ generators and, thus, $m$ massless gauge bosons. The
claim is that we can break part of the symmetry spontaneously, leaving a
subgroup $H$ with $h$ generators unbroken. The $h$ gauge bosons associated to
$H$ remain massless while the $m-h$ others acquire a mass. In order to achieve
this result we need $m-h$ scalar degrees of freedom with the same quantum
numbers as the broken generators. They will disappear from the physical
spectrum and will re-appear as zero helicity states of the massive vector
bosons. As previously, we shall see that one needs at least one more scalar
state which remains physical.

In the remaining of this section we show explicitly these results for a
general gauge group. The reader who is not interested in technical details may
skip this part.

We introduce a multiplet of scalar fields $\phi_i$ which transform according to some representation, not necessarily irreducible, of $G$ of dimension $n$. According to the rules we explained in the last section, the Lagrangian of the system is given by:

\be 
\label{spbrna1}
{\cal L}=-\frac{1}{4}Tr(G_{\mu \nu}G^{\mu \nu}) +(D_{\mu}\Phi)^\dagger D^{\mu}\Phi-V(\Phi)
\ee

In component notation, the covariant derivative is, as usual, $D_{\mu}\phi_i=\partial_{\mu}\phi_i-ig^{(a)}T_{ij}^aA_{\mu}^a\phi_j$ where we have allowed for the possibility of having arbitrary coupling constants $g^{(a)}$ for the various generators of $G$ because we do not assume that $G$ is simple or semi-simple. $V(\Phi)$ is a polynomial in $\Phi$ invariant under $G$ of degree equal to four. As before, we assume that we can choose the parameters in $V$ such that the minimum is not at $\Phi=0$ but rather at $\Phi=v$ where $v$ is a constant vector in the representation space of $\Phi$. $v$ is not unique. The $m$ generators of $G$ can be separated into two classes: $h$ generators which annihilate $v$ and form the Lie algebra of the unbroken subgroup $H$; and $m-h$ generators, represented in the representation of $\Phi$ by matrices $T^a$, such that $T^av\neq0$ and all vectors $T^av$ are independent and can be chosen orthogonal. Any vector in the orbit of $v$, {\it i.e.} of the form $e^{iw^aT^a}v$ is an equivalent minimum of the potential. As before, we should translate the scalar fields $\Phi$ by $\Phi \rightarrow \Phi+v$. It is convenient to decompose $\Phi$ into components along the orbit of $v$ and orthogonal to it, the analogue of the $\chi$ and $\psi$  fields of the previous section. We can write:

\be
\label{spbrna2}
\Phi=i\sum_{a=1}^{m-h}\frac{\chi^aT^av}{|T^av|}+\sum_{b=1}^{n-m+h}\psi^bu^b+v
\ee
where the vectors $u^b$ form an orthonormal basis in the space orthogonal to all $T^av$'s. The corresponding generators span the coset space $G/H$. As before, we shall show that the fields $\chi^a$ will be absorbed by the Higgs mechanism and the fields $\psi^b$ will remain physical. Note that the set of vectors $u^b$ contains at least one element since, for all $a$, we have:

\be
\label{spbrna3}
v\cdot T^av=0
\ee
because the generators in a real unitary representation are
anti-symmetric. This shows that the dimension $n$ of the representation of
$\Phi$ must be larger than $m-h$ and, therefore, there will remain at least
one physical scalar field which, in the quantum theory, will give  a physical
scalar particle\footnote{Obviously, the argument assumes the
  existence of scalar fields which induce the phenomenon of spontaneous
  symmetry breaking. We can construct models in which the role of the
  latter is played by some kind of fermion-antifermion bound states and they come
  under the name of models with a {\it dynamical symmetry breaking}. In such
  models the existence of a physical spin-zero state, the analogue of the
  $\sigma$-particle of the chiral symmetry breaking of QCD,  is a dynamical
  question, in general hard to answer.}.

Let us now bring in the Lagrangian (\ref{spbrna1}) the expression of $\Phi$ from (\ref{spbrna2}). We obtain:

\be
{\cal L} =  \frac{1}{2}\sum_{a=1}^{m-h}(\partial_{\mu}\chi^a)^2+\frac{1}{2}\sum_{b=1}^{n-m+h}(\partial_{\mu}\psi^b)^2-\frac{1}{4}Tr(F_{\mu \nu}F^{\mu \nu})\nonumber
\ee

\be
\label{spbrna4}
+  \frac{1}{2}\sum_{a=1}^{m-h}g^{(a)2}|T^av|^2A_{\mu}^aA^{\mu a} -\sum_{a=1}^{m-h}g^{(a)}T^av\partial^{\mu}\chi^aA_{\mu}^a -V(\Phi)+.....
\ee
where the dots stand for coupling terms between the scalars and the gauge fields. In writing (\ref{spbrna4}) we took into account that $T^bv=0$ for $b>m-h$ and that the vectors $T^av$ are orthogonal.

The analysis that gave us Goldstone's theorem shows that

\be
\label{spbrna5}
\frac{\partial^2V}{\partial\phi_k\partial\phi_l}|_{\Phi=v}(T^av)_l=0
\ee
which shows that the $\chi$-fields would correspond to the Goldstone modes. As a result, the only mass terms which appear in $V$ in equation (\ref{spbrna4}) are of the form $\psi^kM^{kl}\psi^l$ and do not involve the $\chi$-fields. 

As far as the bilinear terms in the fields are concerned, the Lagrangian
(\ref{spbrna4}) is the sum of terms of the form found in the Abelian case. All
gauge bosons which do not correspond to $H$ generators acquire a mass equal to
$m_a=g^{(a)}|T^av|$ and, through their mixing with the would-be Goldstone
fields $\chi$, develop a zero helicity state. All other gauge bosons remain
massless. The $\psi$'s represent the remaining physical Higgs fields.  

\section{Building the STANDARD MODEL: A five step programme}

In this section we shall construct the Standard Model of electro-weak
interactions as a spontaneously broken gauge theory. We shall follow the hints
given by experiment following a five step programme:

$\bullet$ Step 1: Choose a gauge group $G$.

$\bullet$ Step 2: Choose the fields of the ``elementary'' particles and assign them to
representations of 
$G$. Include scalar fields to allow for the Higgs mechanism. 

$\bullet$ Step 3: Write the most general renormalisable Lagrangian invariant
   under $G$. At this stage gauge invariance is still exact and
   all gauge vector bosons are massless.

$\bullet$ Step 4: Choose the parameters of the Higgs potential so that
   spontaneous symmetry breaking occurs.

$\bullet$ Step 5: Translate the scalars and rewrite the Lagrangian in terms of
   the translated fields. Choose a suitable gauge and quantise
   the theory.

{\it A  remark: Gauge theories provide only the general framework,
not a detailed model. The latter will depend on the particular
choices made in steps 1) and 2).}

\subsection{The lepton world}

We start with the leptons and, in order to simplify the presentation, we shall
assume that neutrinos are massless. We follow the five steps:

$\bullet$ Step 1: Looking at the Table of Elementary Particles we see that, for the
combined electromagnetic and weak interactions, we have four gauge bosons, namely
$W^{\pm}$, $Z^0$ and the photon. As we explained earlier, each
one of them corresponds to a generator of the group $G$. The only non-trivial
group with four generators is $U(2)\approx SU(2)\times U(1)$.

Following the notation which was inspired by the hadronic physics, we call
$T_i$, $i=1,2,3$ the three generators of $SU(2)$ and $Y$ that of $U(1)$. Then,
the electric charge operator $Q$ will be a linear combination of $T_3$ and
$Y$. By convention, we write:

\be
\label{charge}
Q=T_3+\frac{1}{2}Y
\ee

The coefficient in front of $Y$ is arbitrary and only fixes the normalisation
of the $U(1)$ generator relatively to those of $SU(2)$\footnote{The
  normalisation of the generators for non-Abelian groups is fixed by their
  commutation relations. That of the Abelian generator is arbitrary. The
  relation (\ref{charge}) is one choice which has only a historical value. It
  is not the most natural one from the group theory point of view, as you will
  see in the discussion concerning Grand-Unified theories.}. This ends our discussion of the first step.

$\bullet$ Step 2: The number and the interaction properties of the gauge
bosons are fixed by the gauge group. This is no more the case with the fermion
fields. In principle, we can choose any number and assign them to any
representation. It follows that the choice here will be dictated by the
phenomenology. 

Leptons have always been considered as elementary
particles. We have six leptons, however, as we noticed already, a  striking
feature of the data is the phenomenon of 
family repetition. We do not understand why Nature chooses to repeat itself
three times, but the simplest way to incorporate this observation to the model
is to use three times the same representations, one for each family. This
leaves $SU(2)$ doublets and/or singlets as the only possible choices. A further
experimental input we shall use is the fact that the charged $W$'s couple only
to the left-handed components of the lepton fields, contrary to the photon
which couples with equal strength to both right and left. These considerations
lead us to assign the left-handed components of the lepton fields to doublets
of $SU(2)$. 

\be
\label{leptrepr1}
\Psi_L^i(x)=\frac{1}{2}(1+\gamma_5)\left( \begin{array}{c}
\nu_i(x)\\{\ell}^-_i(x) \end{array} 
\right)~~~;~~~i=1,2,3
\ee
where we have used the same symbol for the particle and the associated Dirac
field.

The right-handed components are assigned to singlets of $SU(2)$:

\be
\label{leptrepr2} 
\nu_{iR}(x)=\frac{1}{2}(1-\gamma_5)\nu_i(x)~~(?)~~~;~~~{\ell}^-_{iR}(x)=\frac{1}{2}(1-\gamma_5){\ell}^-_i(x)
\ee

The question mark next to the right-handed neutrinos means that the presence
of these fields is not confirmed by the data. We shall drop them in this
lecture, but we may come back to this point later. We shall also simplify the
notation and put ${\ell}^-_{iR}(x)=R_i(x)$. The resulting transformation
properties under local $SU(2)$ transformations are:

\be
\label{su2trans}
\Psi_L^i(x)\rightarrow e^{i\vec{\tau}\vec{\theta}(x)}\Psi_L^i(x)~~~;~~~R_i(x)\rightarrow R_i(x)
\ee
with $\vec{\tau}$ the three Pauli matrices. This assignment and the $Y$ normalisation given by Eq. (\ref{charge}), fix also the $U(1)$ charge and, therefore, the transformation properties of the lepton fields. For all $i$ we find:

\be
\label{leptcharge}
Y(\Psi_L^i)=-1~~~;~~~~Y(R_i)=-2
\ee

If a right-handed neutrino exists, it has $Y(\nu_{iR})=0$, which shows that it
is not coupled to any gauge boson.

We are left with the choice of the Higgs scalar fields and we shall choose the solution with the minimal number
of fields. We must give masses to three vector gauge bosons and keep the
fourth one massless. The latter will be identified with the photon. We recall
that, for every vector boson acquiring mass, a scalar with the same quantum
numbers decouples. At the end we shall remain with at least one physical,
neutral, scalar field. It follows that the minimal number to start with is
four, two charged and two neutral. We choose to put them, under $SU(2)$, into
a complex doublet:

\begin{equation}
\label{hsu2}
\Phi=\left( \begin{array}{c}
\phi^+\\\phi^0
\end{array}
\right) ~~~;~~~\Phi(x) \rightarrow e^{i\vec{\tau}\vec{\theta}(x)}\Phi(x)
\end{equation}
with the conjugate fields $\phi^-$ and ${\phi^{0}}^*$ forming $\Phi^{\dagger}$. The $U(1)$ charge of $\Phi$ is $Y(\Phi)=1$.

This ends our choices for the second step. At this point the model is
complete. All further steps are purely technical and uniquely defined. 

$\bullet$ Step 3:  What follows is straightforward
algebra. We write the most general, renormalisable, Lagrangian, involving the
fields (\ref{leptrepr1}), (\ref{leptrepr2}) and (\ref{hsu2}) invariant under gauge transformations of
$SU(2)\times U(1)$. We shall also assume the separate conservation of the
three lepton numbers, leaving the discussion on the neutrino mixing to a
specialised lecture. The requirement of renormalisability implies that all
terms in the Lagrangian are monomials in the fields and their derivatives and
their canonical dimension is smaller or equal to four. The result is:

\begin{eqnarray}
{\cal {L}} &=&-\frac{1}{4}\vec{W}_{\mu \nu}\cdot \vec{W}^{\mu \nu}-\frac{1}{4}B_{\mu \nu}B^{\mu \nu}+|D_{\mu}\Phi|^2-V(\Phi)\nonumber \\
 &+& \sum_{i=1}^{3}\left [ \bar{\Psi}_L^ii\Dsl \Psi_L^i +\bar{R_i} i\Dsl R_i-G_i(\bar{\Psi}_L^iR_i\Phi +h.c.)\right ]
\label{sml1} 
\end{eqnarray}

If we call $\vec{W}$ and $B$ the gauge fields associated to $SU(2)$ and $U(1)$
respectively, the corresponding field strengths $\vec{W}_{\mu \nu}$ and $B_{\mu \nu}$ appearing in
(\ref{sml1}) are given by (\ref{curv}) and (\ref{Fmunu}).

Similarly, the covariant derivatives in (\ref{sml1}) are determined by the
assumed transformation properties of the fields, as shown in
(\ref{gencovder}):

\begin{equation}
\begin{array}{c} 
D_{\mu}\Psi_L^i=\left( \partial_{\mu} -ig\frac{\vec{\tau}}{2}\cdot
\vec{W}_{\mu}+i\frac{g'}{2}B_{\mu} \right) \Psi_L^i ~~;~~D_{\mu}R_i=\left(
\partial_{\mu}+ig'B_{\mu} \right)R_i\\
D_{\mu}\Phi=\left( \partial_{\mu}-ig\frac{\vec{\tau}}{2}\cdot
\vec{W}_{\mu}-i\frac{g'}{2}B_{\mu} \right)\Phi
\end{array}
\label{sml3}
\end{equation}

The two coupling constants $g$ and $g'$ correspond to the groups $SU(2)$ and $U(1)$
respectively. The most general Higgs potential $V(\Phi)$ compatible with the transformation
properties of the field $\Phi$ is:

\begin{equation}
\label{sml4}
V(\Phi)=\mu^2 \Phi^{\dagger}\Phi+\lambda(\Phi^{\dagger}\Phi)^2
\end{equation}

The last term in (\ref{sml1}) is a Yukawa coupling term between the scalar
$\Phi$ and the fermions. In the absence of  right-handed neutrinos, this is the most general term which is invariant under
$SU(2)\times U(1)$. As usual, $h.c.$ stands for ``hermitian conjugate''. $G_i$
are three  arbitrary coupling constants. If right-handed neutrinos exist there
is a second Yukawa term with $R_i$ replaced by $\nu_{iR}$ and $\Phi$ by the
corresponding doublet proportional to $\tau_2\Phi^*$, where * means ``complex
conjugation''. We see that the 
Standard model can perfectly well accommodate a right-handed neutrino, but it
couples only to the Higgs field.

A final remark: As expected, the gauge bosons $\vec{W}_{\mu}$ and $B_{\mu}$
appear to be massless. The same is true for all fermions. This is not
surprising because the assumed different transformation properties of the right
and left handed components forbid the appearance of a Dirac mass term in the
Lagrangian. On the other hand, the Standard Model quantum numbers also
 forbid the appearance of a
Majorana mass term for the neutrinos. In fact, the only dimensionful parameter in (\ref{sml1}) is
$\mu^2$, the parameter in the Higgs potential (\ref{sml4}). Therefore, the
mass of every particle in the model is expected to be proportional to $|\mu |$.

$\bullet$ Step 4: The next step of our program consists in  choosing the parameter
$\mu^2$ of the Higgs potential negative in order to trigger the phenomenon of
spontaneous symmetry breaking and the Higgs mechanism. The minimum of the
potential occurs at a point $v^2=-\mu^2/\lambda$. As we have explained earlier,
we can choose the direction of the breaking to be along
the real part of $\phi^0$.  

$\bullet$ Step 5: Translating the Higgs field by a real constant:

\begin{equation}
\label{htran}
\Phi \rightarrow \Phi + \frac{1}{\sqrt{2}}\left( \begin{array}{c}
0 \\
v
\end{array}
\right)~~~~~v^2=-\frac{\mu^2}{\lambda}
\end{equation}
transforms the Lagrangian and generates new terms, as it was explained in the
previous section. Let us look at some of them:

{\it (i) Fermion mass terms.} Replacing $\phi^0$ by $v$ in the Yukawa term in
(\ref{sml1}) creates a mass term for the charged leptons, leaving the
neutrinos massless. 

\begin{equation}
\label{lmas}
m_e=\frac{1}{\sqrt{2}}G_e v~~~~m_{\mu}=\frac{1}{\sqrt{2}}G_{\mu} v~~~~m_{\tau}=\frac{1}{\sqrt{2}}G_{\tau} v
\end{equation}

Since we have three arbitrary constants $G_i$, we can fit the three observed
lepton masses. If we introduce right-handed neutrinos we can also fit
whichever Dirac neutrino masses we wish.

{\it (ii) Gauge boson mass terms.} They come from the $|D_{\mu}\Phi|^2$ term
in the Lagrangian. A straight substitution produces the following quadratic
terms among the gauge boson fields:

\begin{equation}
\label{gmas}
\frac{1}{8}v^2[g^2(W^1_{\mu}W^{1\mu}+W^2_{\mu}W^{2\mu})+(g'B_{\mu}-gW^3_{\mu})^2]
\end{equation}

Defining the charged vector bosons as:

\begin{equation}
\label{cvb}
W^{\pm}_{\mu}=\frac{W^1_{\mu}\mp iW^2_{\mu}}{\sqrt2}
\end{equation}
we obtain their masses:

\begin{equation}
\label{cgmas}
m_W=\frac{vg}{2}
\end{equation}

The neutral gauge bosons $B_{\mu}$ and $W^3_{\mu}$ have a 2$\times$2
non-diagonal mass matrix. After diagonalisation, we define the mass
eigenstates:

\begin{equation}
\label{phz}
\begin{split}
Z_{\mu} &=\cos\theta_W B_{\mu} -\sin\theta_W W^3_{\mu}\\
A_{\mu} &=\cos\theta_W B_{\mu}+ \sin\theta_W W^3_{\mu}
\end{split}
\end{equation}
with $\tan\theta_W=g'/g$. They correspond to the mass eigenvalues

\begin{equation}
\label{ngmas}
\begin{split}
m_Z &=\frac{v(g^2+{g'}^2)^{1/2}}{2}=\frac{m_W}{\cos\theta_W}\\
m_A &=0
\end{split}
\end{equation}

As expected, one of the neutral gauge bosons is massless and will be identified
with the photon. The Higgs mechanism breaks the original symmetry according to $SU(2)\times U(1)\rightarrow U(1)_{em}$ and $\theta_W$ is the angle between the original $U(1)$ and the one left unbroken. It is the parameter first introduced by S.L. Glashow, although it is often referred to as ``Weinberg angle''. 

{\it (iii) Physical Higgs mass.} Three out of the four real fields of the
$\Phi$ doublet will be absorbed by the Higgs mechanism in order to allow for
the three gauge bosons $W^{\pm}$ and $Z^0$ to acquire a mass. The fourth one,
which corresponds to $(|\phi^0 \phi^{0\dagger}|)^{1/2}$, remains physical. Its
mass is given by the coefficient of the quadratic part of $V(\Phi)$ after the
translation (\ref{htran}) and is equal to:

\begin{equation}
\label{hmas}
m_H=\sqrt{-2\mu^2}=\sqrt{2\lambda v^2}
\end{equation}

In addition, we produce various coupling terms which we shall present,
together with the hadronic ones, in the next section.

\subsection{Extension to hadrons}

Introducing the hadrons into the model presents some novel features. They are
mainly due to the fact that the individual quark quantum numbers are not
separately conserved. As regards to the second step,  today there is a
consensus regarding the choice of the  
``elementary'' constituents of matter: Besides the six leptons, there are six
quarks. They are fractionally charged and come each in three
``colours''. The observed lepton-hadron
universality property, tells us to use also doublets and singlets for the
quarks. The first novel feature we mentioned above is that all quarks appear
to have non-vanishing Dirac masses, so we must introduce both right-handed
singlets for each family. A na\"{i}ve assignment would be to write the analogue
of Equations (\ref{leptrepr1}) and (\ref{leptrepr2}) as: 

\be
\label{quarkrepr1}
Q_L^i(x)=\frac{1}{2}(1+\gamma_5)\left( \begin{array}{c}
U^i(x)\\D^i(x) \end{array} 
\right)~~~;~~~U^i_R(x)~~~;~~~D^i_R(x)
\ee
with the index $i$ running over the three families as $U^i=u,c,t$ and
$D^i=d,s,b$ for $i=1,2,3$, respectively\footnote{An additional index $a$, running also through 1,2 and 3 and denoting the colour, is understood.}. This assignment determines the
$SU(2)$ transformation properties of the quark fields. It also fixes
their $Y$ charges and, hence their $U(1)$ properties. Using
Eq. (\ref{charge}), we find 

\be
\label{quarkY}
Y(Q_L^i)=\frac{1}{3}~~~;~~~Y(U^i_R)=\frac{4}{3}~~~;~~~Y(D^i_R)=-\frac{2}{3}
\ee
 
The presence of the two right handed singlets has an important consequence. Even if we had only one
family, we would have two distinct
Yukawa terms between the quarks and the Higgs field of the form:

\begin{equation}
\label{smh1} 
{\cal {L}}_{Yuk}=G_d(\bar{Q}_L D_R \Phi +h.c.)+G_u(\bar{Q}_L U_R \tilde{\Phi} +h.c.)
\end{equation}

$\tilde{\Phi}$ is the doublet proportional to $\tau_2\Phi^*$. It has the
same transformation properties under $SU(2)$ as $\Phi$, but the opposite $Y$
charge.

If there were only one family, this would have been the end of the story. The hadron Lagrangian
${\cal {L}}_h^{(1)}$ is the same as (\ref{sml1}) with quark fields replacing
leptons and the extra term of
(\ref{smh1}). The complication we
alluded to before comes with the addition of more families. In this case the
total Lagrangian is not just the sum  over the family
index.  The physical reason is the non-conservation of the individual quark
quantum numbers we mentioned previously. In writing (\ref{quarkrepr1}), we  implicitly
assumed a particular pairing of the quarks in each family, $u$ with $d$, $c$
with $s$ and $t$ with $b$. In general, we could choose any  basis in family
space and, since we have two 
 Yukawa terms, we will not be able to diagonalise both of them
 simultaneously. It follows that 
the most general Lagrangian will contain a matrix with non-diagonal terms
which mix the families. By convention, we attribute it to a different choice
of basis in the $d-s-b$ space. It follows that the correct generalisation of
the Yukawa Lagrangian (\ref{smh1}) to many families is given by:

\begin{equation}
\label{smh11} 
{\cal {L}}_{Yuk}=\sum_{i,j} \left [ (\bar{Q}_L^iG_d^{ij} D_R^j \Phi
  +h.c.)\right ]+\sum_{i} \left [G_u^i(\bar{Q}_L^i U_R^i \tilde{\Phi} +h.c.)\right ]
\end{equation}
where the Yukawa coupling constant $G_d$ has become a matrix in family
space. After translation of the Higgs field, we shall produce masses for the
up quarks given by $m_u=G_u^1v$, $m_c=G_u^2v$ and $m_t=G_u^3v$, as well as a three-by-three
mass matrix for the down quarks given by $G_d^{ij}v$. As usually, we want to work in a field space
where the masses are diagonal, so we change our initial $d-s-b$ basis to bring
$G_d^{ij}$ into a diagonal form. This can be done through a three-by-three
unitary matrix $\tilde{D}^i=U^{ij}D^j$ such that $U^\dagger G_d U=$
diag$(m_d,m_s,m_b)$ .  
In the simplest example of only two families, it is
easy to show that the most general such matrix, after using all freedom for
field redefinitions and phase choices,  is a real rotation: 

\begin{equation}
\label{cab}
C=\left( \begin{array}{cc} 
\cos\theta & \sin\theta \\
- \sin\theta & \cos\theta
\end{array}
\right)
\end{equation}
with $\theta$ being our familiar Cabibbo angle. For three families an easy
counting shows that the matrix has three angles, the three Euler angles,
and an arbitrary
phase. It is traditionally written in the form:

\begin{equation}
\label{km}
KM=\left( \begin{array}{ccc} 
c_1 & s_1 c_3 & s_1 s_3\\
-s_1 c_3 & c_1 c_2 c_3 -s_2s_3e^{i\delta} & c_1 c_2 s_3 +s_2 c_3e^{i\delta} \\
-s_1 s_2 & c_1 s_2 c_3 +c_2 s_3e^{i\delta}  & c_1 s_2 s_3 -c_2 c_3e^{i\delta}
\end{array}
\right)
\end{equation}
with the notation $c_k=\cos\theta_k$ and $s_k=\sin\theta_k$, $k=1,2,3$. The novel feature
is the possibility of introducing  the phase $\delta$. This means that a
  six-quark model has a natural source of $CP$, or $T$, violation, while a four-quark
  model does not. 

The total Lagrangian density, before the translation of the Higgs field, is
now:

\begin{eqnarray}
\label{totWS}
{\cal {L}} &=&-\frac{1}{4}\vec{W}_{\mu \nu}\cdot \vec{W}^{\mu \nu}-\frac{1}{4}B_{\mu \nu}B^{\mu \nu}+|D_{\mu}\Phi|^2-V(\Phi)\nonumber \\
 &+& \sum_{i=1}^{3}\left [ \bar{\Psi}_L^ii\Dsl \Psi_L^i +\bar{R_i} i\Dsl
 R_i-G_i(\bar{\Psi}_L^iR_i\Phi +h.c.) \right. \\ \nonumber
 &+& \left. \bar{Q}_L^ii\Dsl Q_L^i +\bar{U}_R^ii\Dsl U_R^i + \bar{D}_R^ii\Dsl
 D_R^i +G_u^i(\bar{Q}_L^i U_R^i
 \tilde{\Phi} +h.c.) \right ]\\ \nonumber
 &+& \sum_{i,j=1}^{3}\left [ (\bar{Q}_L^iG_d^{ij} D_R^j \Phi +h.c.) \right ] 
\end{eqnarray}

The covariant derivatives on the quark fields are given by:

\begin{eqnarray}
\label{smh2} 
D_{\mu}Q_L^i=\left( \partial_{\mu} -ig\frac{\vec{\tau}}{2}\cdot
\vec{W}_{\mu}-i\frac{g'}{6}B_{\mu} \right) Q_L^i\\ \nonumber
 D_{\mu}U_R^i=\left(
\partial_{\mu}-i\frac{2g'}{3}B_{\mu} \right)U_R^i\\ \nonumber 
D_{\mu}D_R^i=\left( \partial_{\mu}+i\frac{g'}{3}B_{\mu} \right)D_R^i
\end{eqnarray} 

The classical Lagrangian (\ref{totWS})
contains seventeen arbitrary real parameters. They are:

-The two gauge coupling constants $g$ and $g'$. 

-The two parameters of the Higgs potential $\lambda$ and $\mu^2$.

-Three Yukawa coupling constants for the three lepton families, $G_{e,\mu ,\tau}$.

-Six Yukawa coupling constants for the three quark families, $G_u^{u,c,t}$,
 and $G_d^{d,s,b}$.

-Four parameters of the $KM$ matrix, the three angles and the phase $\delta$.

A final remark: Fifteen out of these seventeen parameters are  directly
connected with the Higgs sector.

Translating the Higgs field by Eq. (\ref{htran}) and diagonalising the
resulting down quark mass matrix produces the mass terms for fermions and
bosons which we introduced before as well as several coupling terms. We shall
write here the ones which involve the physical fields\footnote{We know from
  quantum electrodynamics that, in order to determine the Feynman rules of a
  gauge theory, one must first decide on a choice of gauge. For Yang-Mills
  theories this step introduces new fields called {\it Faddeev-Popov
    ghosts}. This point is explained in every standard text book on quantum
  field theory, but we  have not discussed it in these lectures.}.

{\it (i) The gauge boson fermion couplings.} They are the ones which generate
the known weak and electromagnetic interactions. $A_{\mu}$ is coupled to the charged fermions through the usual electromagnetic current.

\begin{equation}
\label{emc}
\frac{gg'}{(g^2+{g'}^2)^{1/2}}\left[ \bar{e}\gamma^{\mu}e+\sum_{a=1}^{3}\left(\frac{2}{3}\bar{u}^a\gamma^{\mu}u^a-\frac{1}{3}\bar{d}^a\gamma^{\mu}d^a\right) ~~+...\right]A_{\mu} 
\end{equation}
where the dots stand for the contribution of the other two families $e
\rightarrow \mu ,\tau$, $u \rightarrow c,t$ and $d \rightarrow s,b$ and the
summation over $a$ extends over the three colours. Equation (\ref{emc}) shows that the
electric charge $e$ is given, in terms of $g$ and $g'$ by

\begin{equation}
\label{elc}
e=\frac{gg'}{(g^2+{g'}^2)^{1/2}}=g\sin\theta_W=g'\cos\theta_W
\end{equation}

Similarly, the couplings of the charged $W$'s to the weak current are:

\begin{equation}
\label{cwc}
\frac{g}{2\sqrt{2}}\left( \bar{\nu}_e\gamma^{\mu}(1+\gamma_5)e+\sum_{a=1}^{3}\bar{u}^a\gamma^{\mu}(1+\gamma_5)d^a_{KM}~~+...\right)W^+_{\mu}~~+h.c.
\end{equation}

Combining all these relations, we can determine the experimental value of the
parameter $v$, the vacuum expectation value of the Higgs field. We find $v\sim
246$GeV. 

As expected, only left-handed fermions participate. $d_{KM}$ is the linear
combination of $d-s-b$ given by the $KM$ matrix (\ref{km}). By diagonalising
the down quark mass matrix we introduced the off-diagonal terms into the
hadron current. When considering processes, like nuclear $\beta$-decay, or
$\mu$-decay, where the momentum transfer is very small compared to the $W$
mass, the $W$ propagator can be approximated by ${m_W}^{-2}$ and the effective
Fermi coupling constant is given by:

\begin{equation}
\label{Fc}
\frac{G}{\sqrt{2}}=\frac{g^2}{8m_W^2}=\frac{1}{2v^2} 
\end{equation}

Contrary to the charged weak current (\ref{cwc}), the $Z^0$-fermion couplings involve both left- and right-handed fermions:  

\begin{equation}
\label{lnc}
\begin{split}
-\frac{e}{2}\frac{1}{\sin\theta_W \cos\theta_W} & \left[\bar{\nu}_L \gamma^{\mu}\nu_L +
 (\sin^2\theta_W-\cos^2\theta_W)\bar{e}_L \gamma^{\mu} e_L\right.\\ 
 &\left. +2\sin^2\theta_W\bar{e}_R \gamma^{\mu} e_R ~+~...\right]Z_{\mu}
\end{split}
\end{equation}  

\begin{equation}
\label{qnc}
\begin{split}
\frac{e}{2}\sum_{a=1}^{3} &
\left[(\frac{1}{3}\tan\theta_W-\cot\theta_W)\bar{u}^a_L\gamma^{\mu}u^a_L+(\frac{1}{3}\tan\theta_W+\cot\theta_W)\bar{d}^a_L\gamma^{\mu}d^a_L\right.\\
 &
  \left. +\frac{2}{3}\tan\theta_W(2\bar{u}^a_R\gamma^{\mu}u^a_R-\bar{d}^a_R\gamma^{\mu}d^a_R) ~+~...\right]Z_{\mu}      
\end{split}
\end{equation}

Again, the summation is over the colour indices and the dots stand for the contribution of the other two families. We verify in this formula the property of the weak neutral current to be
diagonal in the quark flavour space. Another interesting property is that the
axial part of the neutral current is proportional to $[\bar
  {u}\gamma_{\mu}\gamma_5u-\bar {d}\gamma_{\mu}\gamma_5d]$. This particular
form of the coupling is important for the phenomenological applications, such
as the induced parity violating effects in atoms and nuclei.  

{\it (ii) The gauge boson self-couplings.} One of the characteristic features
of Yang-Mills theories is the particular form of the self couplings among the
gauge bosons. They come from the square of the non-Abelian curvature in the Lagrangian,
which, in our case, is the term $-\frac{1}{4}\vec{W}_{\mu \nu}\cdot
\vec{W}^{\mu \nu}$. Expressed in terms of the physical fields, this term
gives:

\begin{equation}
\label{gbsc}
\begin{split}
 & -ig (\sin\theta_WA^{\mu}-\cos\theta_WZ^{\mu})(W^{\nu -}W_{\mu
 \nu}^+-W^{\nu +}W_{\mu \nu}^-)\\
 & -ig(\sin\theta_WF^{\mu \nu}-\cos\theta_WZ^{\mu
 \nu})W_{\mu}^-W_{\nu}^+\\
 & -g^2( \sin\theta_WA^{\mu}-\cos\theta_WZ^{\mu})^2W_{\nu}^+W^{\nu -}\\
 & +g^2(
 \sin\theta_WA^{\mu}-\cos\theta_WZ^{\mu})(
 \sin\theta_WA^{\nu}-\cos\theta_WZ^{\nu})W_{\mu}^+W_{\nu}^- \\
 & -\frac{g^2}{2}(W_{\mu}^+W^{\mu -})^2+\frac{g^2}{2}(W_{\mu}^+W_{\nu}^-)^2
\end{split}
\end{equation} 
where we have used the following notation: $F_{\mu \nu}=\partial_{\mu}A_{\nu}-\partial_{\nu}A_{\mu}$, $W_{\mu
  \nu}^{\pm}= \partial_{\mu}W_{\nu}^{\pm}-\partial_{\nu}W_{\mu}^{\pm}$ and
  $Z_{\mu \nu}=\partial_{\mu}Z_{\nu}-\partial_{\nu}Z_{\mu}$ with
  $g\sin\theta_W=e$. Let us concentrate on the photon-$W^+ W^-$ couplings. If
  we forget, for the moment, about the $SU(2)$ gauge  
  invariance, we can use different coupling constants for the two trilinear
  couplings in (\ref{gbsc}), say $e$ for the first and $e\kappa$ for the
  second. For a charged, massive $W$, the magnetic moment $\mu$ 
  and the
  quadrupole moment $Q$ are given by:
\begin{equation}
\label{gbempar}
\mu=\frac{(1+\kappa)e}{2m_W}~~~~~Q=-\frac{e\kappa}{m_W^2}
\end{equation}

Looking at (\ref{gbsc}), we see that $\kappa=1$. Therefore,
 $SU(2)$ gauge invariance gives very specific predictions concerning the
electromagnetic parameters of the charged vector bosons. The gyromagnetic
 ratio equals two and the quadrupole moment equals $-e m_W^{-2}$. 

{\it (iii) The scalar Higgs fermion couplings.} They are given by the Yukawa
terms in (\ref{sml1}). The same couplings generate the fermion masses through
spontaneous symmetry breaking. It follows that the physical Higgs scalar
couples to quarks and leptons with strength proportional to the fermion
mass. Therefore the prediction is that it will decay predominantly to the
heaviest possible fermion compatible with phase space. This property provides
a typical signature for Higgs identification.

{\it (iv) The scalar Higgs gauge boson couplings.} They come from the
covariant  derivative term $|D_{\mu}\Phi|^2$ in the Lagrangian. If we call
$\phi$ the field of the physical neutral Higgs, we find: 

\be
\label{gbhc}
\frac{1}{4}(v+\phi)^2\left [ g^2W_{\mu}^+W^{-\mu}+(g^2+g'^2)Z_{\mu}Z^{\mu}\right ]
\ee 

This gives a direct coupling $\phi-W^+-W^-$, as well as $\phi-Z-Z$, which has
been very useful in the Higgs searches.

{\it (v) The scalar Higgs self couplings.} They are proportional to $\lambda
(v+\phi)^4$. Equations (\ref{hmas}) and (\ref{Fc}) show that
$\lambda=Gm_H^2/\sqrt{2}$, so, in the tree approximation, this coupling is related to the Higgs mass.  It
could provide a test of the Standard Model Higgs, but it will not be easy to
measure. On the other hand this relation shows that, if the physical Higgs is
very heavy, it is also strongly interacting and this sector of the model
becomes non-perturbative. 

The five step program is now complete for both leptons and quarks.  The
seventeen parameters of the model 
have all been  determined by experiment. Although the number of arbitrary
parameters seems very large, we should not forget that they are all mass and
coupling parameters, like the electron mass and the fine structure constant of
quantum electrodynamics. The reason we have more of them is that the Standard
Model describes in a unified framework a much larger number of particles and
interactions. 

\section{The Standard Model and Experiment}

Our confidence in this model is amply justified on the basis of its ability to accurately describe the bulk of our present-day 
data and, especially, of its enormous success in predicting new phenomena. Let
us mention a few of them. We shall follow the historical order.

$\bullet$ The discovery of weak neutral currents by Gargamelle in 1972
\vskip 0.2cm
$\nu_{\mu}+e^-\rightarrow \nu_{\mu}+e^-~~~;~~~\nu_{\mu}+N \rightarrow
\nu_{\mu}+X$
\vskip 0.2cm
Both, their strength and their properties were predicted by the Model.
\vskip 0.2cm
$\bullet$ The discovery of charmed particles at SLAC in 1974
\vskip 0.2cm
Their presence was essential to ensure the absence of strangeness changing
neutral currents, ex. $K^0\rightarrow \mu^++\mu^-$
\vskip 0.2cm
Their characteristic property is to decay predominantly in strange particles.
\vskip 0.2cm
$\bullet$ A necessary condition for the consistency of the Model is that $\sum_i Q_i=0$
inside each family.
\vskip 0.2cm
When the $\tau$ lepton was discovered this implied a prediction for the
existence of the $b$ and $t$ quarks 
with the right electric charges. 
\vskip 0.2cm
$\bullet$ The discovery of the $W$ and $Z$ bosons at CERN in 1983 with the
masses predicted by the theory.
\vskip 0.2cm
The characteristic relation of the Standard Model with an isodoublet Higgs
mechanism $m_Z=m_W/\cos\theta_W$ has been checked with very high accuracy
(including 
radiative corrections).
\vskip 0.2cm
$\bullet$ The $t$-quark was {\it seen} at LEP through its effects in radiative
corrections before its actual discovery at Fermilab. 
\vskip 0.2cm
$\bullet$ The vector boson self-couplings, $\gamma-W^+-W^-$ and $Z^0-W^+-W^-$
have been measured at LEP and confirm the Yang-Mills predictions given in
equation (\ref{gbempar})
\vskip 0.2cm
$\bullet$ The recent discovery of a new boson which could be the Higgs
particle of the Standard Model is the last of this impressive series of
successes.
\vskip 0.2cm
All these discoveries should not make us forget that the Standard Model has
been equally successful in fitting a large number of experimental results. You
have all seen the global fit given in Figure \nolinebreak \ref{globfit}. The conclusion is
obvious: {\it The Standard Model has been enormously successful.}

\begin{figure}
\centering
\epsfxsize=6cm
\epsffile{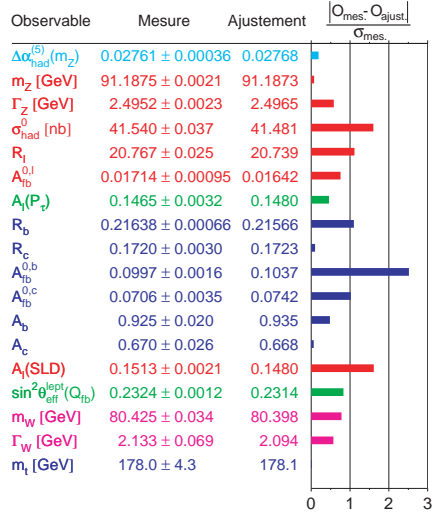}
\caption {A comparison between measured and computed values for various
  physical quantities} \label{globfit} 
\end{figure}

Although in these lectures we did not discuss quantum chromodynamics, the
gauge theory of strong interactions, the computations whose results are
presented in Figure \ref{globfit}, take into account the radiative corrections
induced by virtual gluon exchanges. The fundamental property of quantum
chromodynamics, the one which allows for perturbation theory calculations, is
the property of asymptotic freedom, the particular dependence of the effective
coupling strength on the energy scale. This is presented in Figure \nolinebreak\ref{alphas}. 
The green region shows the theoretical prediction based on QCD
calculations, including the theoretical uncertainties. We see that the
agreement with the experimentally measured values of the effective strong
interaction coupling constant $\alpha_s$ is truly remarkable.  Notice also
that this agreement extends to rather low values of $Q$ of the order of 1-2
GeV, where $\alpha_s$ equals approximately 1/3.

\begin{figure}
\centering
\epsfxsize=6cm
\epsffile{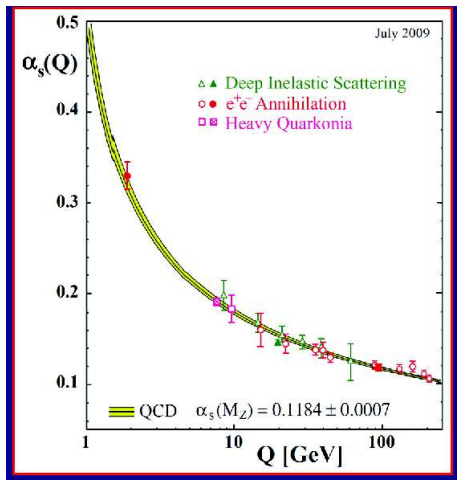}
\caption {The effective coupling constant for strong interactions as a
  function of the energy scale.} \label{alphas} 
\end{figure}

This brings us to our next point, namely that all this success is in fact {\it
  a success of renormalised perturbation theory.} The extreme accuracy of the
experimental measurements, mainly at LEP, but also at FermiLab and elsewhere,
allow, for the first time to make a detailed comparison between theory and
experiment {\it including the purely weak interaction radiative corrections.} 

\begin{figure}
\centering
\epsfxsize=13cm
\epsffile{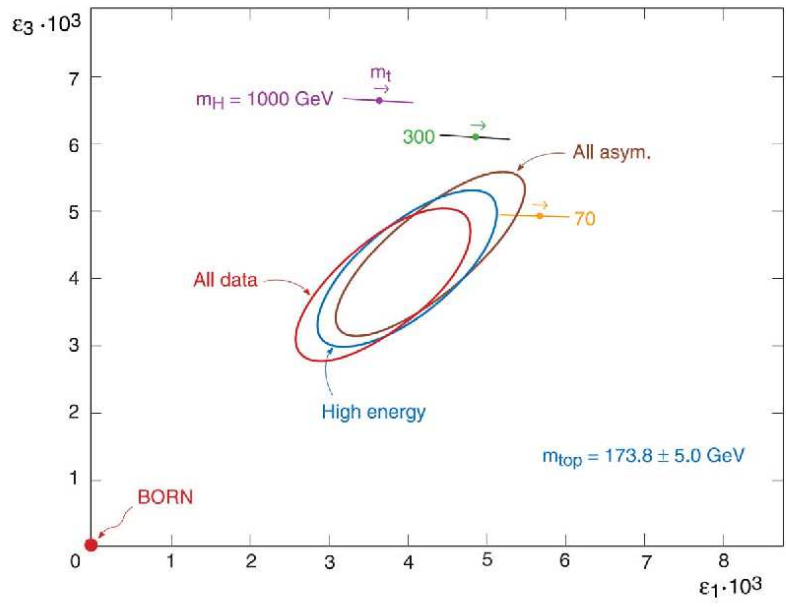}
\caption {Comparison between theory and experiment for two quantities
  sensitive to weak interaction radiative corrections.} \label{epsilons} 
\end{figure}

In Figure \ref{epsilons} we show the comparison between theory and experiment
for two quantities, $\epsilon_1$ and $\epsilon_3$, defined in equations
(\ref{epsilon_1}) and (\ref{epsilon_3}), respectively:

\be
\label{epsilon_1}
\epsilon_1=\frac{3G_Fm_t^2}{8\sqrt{2}\pi^2}-\frac{3G_Fm_W^2}{4\sqrt{2}\pi^2}\tan^2\theta_W \ln \frac{m_H}{m_Z}+...
\end{equation}

\be
\label{epsilon_3}
\epsilon_3=\frac{G_Fm_W^2}{12\sqrt{2}\pi^2}\ln \frac{m_H}{m_Z}-\frac{G_Fm_W^2}{6\sqrt{2}\pi^2} \ln \frac{m_t}{m_Z}+...
\end{equation}

They are  defined with the following properties: (i) They include the strong
and electromagnetic radiative corrections and (ii), they vanish in the Born
approximation for the weak interactions. So, they measure the purely weak
interaction radiative corrections. The Figure is based on a fit which is
rather old and does not include 
the latest data but, nevertheless, it shows that, in order to obtain
agreement with the data, one must include these
corrections. Weak interactions are no more a simple phenomenological model,
but have become a precision theory. 

 The moral of the story is that the perturbation expansion of the Standard
Model is reliable as long as all coupling constants remain small. The only
coupling which does become large in some kinematical regions is $\alpha_s$
which grows at small energy scales, as shown in Figure \ref{alphas}. In
this region we know that a hadronisation process occurs and perturbation
theory breaks down. We
conclude that at high energies perturbation theory is expected to be reliable
{\it unless there are new strong interactions.}

This brings us to our last point, namely 
that this very success shows also that the Standard Model
cannot be a complete theory, in other words there must be new Physics beyond
the Standard Model. The argument is simple and it is based on a
straightforward application of perturbation theory with an additional
assumption which we shall explain presently.

We assume that the Standard Model is correct up to a certain scale
$\Lambda$. The precise value of $\Lambda$ does not matter, provided it is 
larger than any energy scale reached so far\footnote{The scale $\Lambda$ should not be confused with a cut-off
  one often introduces when computing Feynman
  diagrams. This cut-off disappears after renormalisation is performed. Here
  $\Lambda$ is a physical scale which indicates how far the theory can be
  trusted.}.

A quantum field theory is defined through a functional integral over all
classical field
configurations, the Feynman path integral. By a Fourier transformation we can
express it as an integral over the fields defined in momentum space. Following
K. Wilson, let us split this integral in two parts: the high energy part with
modes above $\Lambda$ and the low energy part with the modes below
$\Lambda$. Let us imagine that we perform the high energy part. The result
will be an effective theory expressed in terms of the low energy modes of the
fields. We do
not know how to perform this integration explicitly, so we cannot write down
the correct low energy theory, but the most general form will be a series of
operators made out of powers of the fields and their derivatives. Since
integrating over the heavy modes does not break any of the symmetries of the
initial Lagrangian, only operators allowed by the symmetries will appear. Wilson
remarked that, when $\Lambda$ is large compared to the mass parameters of the
theory,  we can determine the leading contributions by simple dimensional
analysis\footnote{There are some additional technical assumptions concerning
  the dimensions of the fields, but they are satisfied in perturbation
  theory.}. We distinguish three kinds of operators, according to their
canonical dimension: 

$\bullet$ Those with dimension larger than four. Dimensional analysis shows
that they will come with a
coefficient proportional to inverse powers of $\Lambda$, so, by choosing the
scale large enough, we can make their contribution arbitrarily small. We shall
call them {\it irrelevant operators}.

$\bullet$ Those with dimension equal to four. They are the ones which appeared
already in the original Lagrangian. Their coefficient will be independent of
$\Lambda$, up to logarithmic corrections which we ignore. We shall
call them {\it marginal operators}.

$\bullet$ Finally we have the operators with dimension smaller than four. In
the Standard Model there is only one such operator, the square of the Higgs
field $\Phi^2$ which has dimension equal to two\footnote{One could think of
  the square of a fermion operator $\bar{\Psi}\Psi$, whose dimension is equal
  to three, but it is not allowed by
  the chiral symmetry of the model.}. This operator will appear with a
coefficient proportional to $\Lambda^2$, which means that its contribution
will grow quadratically with $\Lambda$. We shall call it {\it relevant
  operator}. It will give an effective mass to the scalar field proportional
to the square of whichever scale we can think of. This problem was first
identified in the framework of Grand Unified Theories and is known since as
{\it the hierarchy problem}. Let me emphasise here that this does not mean
that the mass of the scalar particle will be necessarily equal to
$\Lambda$. The Standard Model is a renormalisable theory and the mass is fixed
by a renormalisation condition to its physical value. It only means that this
condition should be adjusted to arbitrary precision order by order in
perturbation theory. It is this extreme sensitivity to high scales, known as
{\it the fine tuning problem}, which is considered unacceptable for a
fundamental theory. 

Let us summarise: The great success of the Standard Model tells us that
renormalised perturbation theory is reliable in the absence of strong
interactions. The same perturbation theory shows the need of a fine tuning for
the mass of the scalar particle. If we do not accept the latter, we have the following
two options:

$\bullet$ Perturbation theory breaks down at some scale $\Lambda$. We can
imagine several reasons for a such a breakdown to occur. The simplest is the
appearance of new strong interactions. The so called {\it Technicolor} models,
in which the role of the Higgs field is played by a bound state of new strongly coupled
fermions, were in this class. More exotic possibilities include the appearance
of new, compact space dimensions with compactification length $\sim
\Lambda^{-1}$.

$\bullet$ Perturbation theory is still valid but the numerical coefficient  of
the $\Lambda^2$ term which multiplies the $\Phi^2$ operator vanishes to all
orders of perturbation theory. For this to happen we must modify the Standard
Model introducing appropriate new particles. Supersymmetry is the only
systematic way we know to achieve this goal.

\section{Conclusions}

In these lectures we saw the fundamental role of Geometry in the Dynamics of
the forces among the elementary particles. It was the understanding of this role which
revolutionised our way of thinking and led to the construction of the Standard
Model. It incorporates the ideas of gauge theories, as well as those of
spontaneous symmetry breaking. Its agreement with experiment is
spectacular. It fits all data known today. However, 
unless one is willing to accept a fine tuning with
arbitrary precision, one should conclude that New Physics will appear beyond
a scale $\Lambda$. The precise value of $\Lambda$ cannot be computed, but the
amount of fine tuning grows quadratically with it, so it cannot be too
large. Hopefully, it will be within reach of the LHC. 

\section{Acknowledgments}

I wish to thank all the participants to the 2012 CERN Summer School who, by their
questions and remarks, helped me in sharpening the arguments presented in
these notes and, in particular, Dr Christophe Grojean for a critical reading
of the manuscript.

\end{document}